\documentclass[preprint2]{aastex63}
\usepackage{color}
\usepackage{newtxtext,newtxmath}

\usepackage[T1]{fontenc}
\usepackage{ae,aecompl}
\usepackage{longtable}

\usepackage[english]{babel}

\usepackage{graphicx}

\newcommand{\name}{ZTF18abvkwla}
\newcommand{\swift}{\emph{Swift}}

\newcommand{\code}[1]{\texttt{\detokenize{#1}}}
\newcommand{\ct}{\mbox{$\rm count$}}
\newcommand{\tfade}{\mbox{$\rm t_{fade}$}}
\newcommand{\trise}{\mbox{$\rm t_{rise}$}}

\newcommand{\mni}{\mbox{$\rm M_{Ni}$}}

\newcommand{\erg}{\mbox{$\rm erg$}}
\newcommand{\psec}{\mbox{$\rm s^{-1}$}}
\newcommand{\phz}{\mbox{$\rm Hz^{-1}$}}
\newcommand{\cm}{\mbox{$\rm cm$}}
\newcommand{\kev}{\mbox{$\rm keV$}}
\newcommand{\pcmsq}{\mbox{$\rm cm^{-2}$}}
\newcommand{\pcmcub}{\mbox{$\rm cm^{-3}$}}

\newcommand{\pmpccub}{\mbox{$\rm Mpc^{-3}$}}

\newcommand{\ujy}{\mbox{$\rm \mu Jy$}}
\newcommand{\km}{\mbox{$\rm km$}}
\newcommand{\kpc}{\mbox{$\rm kpc$}}
\newcommand{\ghz}{\mbox{$\rm GHz$}}
\newcommand{\days}{\mbox{$\rm d$}}
\newcommand{\msol}{\mbox{$M_\odot$}}
\newcommand{\pyr}{\mbox{$\rm yr^{-1}$}}
\newcommand{\pgyr}{\mbox{$\rm Gyr^{-1}$}}
\newcommand{\yr}{\mbox{$\rm yr$}}
\newcommand{\nickel}{\mbox{$\rm {}^{56}Ni$}}
\newcommand{\degsq}{\mbox{$\rm deg^{2}$}}

\newcommand{\NII}{[\mbox{N\hspace{0.15em}{\sc ii}}]}
\newcommand{\OI}{\mbox{O\hspace{0.15em}{\sc i}}}
\newcommand{\OII}{[\mbox{O\hspace{0.15em}{\sc ii}}]}
\newcommand{\OIII}{[\mbox{O\hspace{0.15em}{\sc iii}}]}
\newcommand{\SII}{[\mbox{S\hspace{0.15em}{\sc ii}}]}
\newcommand{\NeIII}{[\mbox{Ne\hspace{0.15em}{\sc iii}}]}
\newcommand{\HeI}{\mbox{He\hspace{0.15em}{\sc i}}}
\newcommand{\ArIII}{[\mbox{Ar\hspace{0.15em}{\sc iii}}]}

\begin{document}

\title{
The Koala: A Fast Blue Optical Transient with Luminous Radio Emission from a Starburst Dwarf Galaxy at $z=0.27$
}

\author[0000-0002-9017-3567]{Anna Y. Q.~Ho}
\affiliation{Cahill Center for Astrophysics, 
California Institute of Technology, MC 249-17, 
1200 E California Boulevard, Pasadena, CA, 91125, USA}

\author[0000-0001-8472-1996]{Daniel A.~Perley}
\affiliation{Astrophysics Research Institute, Liverpool John Moores University, IC2, Liverpool Science Park, 146 Brownlow Hill, Liverpool L3 5RF, UK}

\author{S. R.~Kulkarni}
\affiliation{Cahill Center for Astrophysics, 
California Institute of Technology, MC 249-17, 
1200 E California Boulevard, Pasadena, CA, 91125, USA}

\author{Dillon Z. J. Dong}
\affiliation{Cahill Center for Astrophysics, 
California Institute of Technology, MC 249-17, 
1200 E California Boulevard, Pasadena, CA, 91125, USA}

\author{Kishalay De}
\affiliation{Cahill Center for Astrophysics, 
California Institute of Technology, MC 249-17, 
1200 E California Boulevard, Pasadena, CA, 91125, USA}

\author[0000-0002-0844-6563]{Poonam Chandra}
\affiliation{National Centre for Radio Astrophysics, Tata Institute of Fundamental Research, PO Box 3, Pune,
411007, India}

\author{Igor Andreoni}
\affiliation{Cahill Center for Astrophysics, 
California Institute of Technology, MC 249-17, 
1200 E California Boulevard, Pasadena, CA, 91125, USA}

\author[0000-0001-8018-5348]{Eric C. Bellm}
\affiliation{DIRAC Institute, Department of Astronomy, University of Washington, 3910 15th Avenue NE, Seattle, WA 98195, USA}

\author{Kevin B. Burdge}
\affiliation{Cahill Center for Astrophysics, 
California Institute of Technology, MC 249-17, 
1200 E California Boulevard, Pasadena, CA, 91125, USA}

\author{Michael Coughlin}
\affiliation{School of Physics and Astronomy, University of Minnesota,
Minneapolis, Minnesota 55455, USA}

\author{Richard Dekany}
\affiliation{Caltech Optical Observatories, California Institute of Technology, Pasadena, CA  91125}

\author{Michael Feeney}
\affiliation{Caltech Optical Observatories, California Institute of Technology, Pasadena, CA  91125}

\author[0000-0002-1153-6340]{Dmitry D. Frederiks}
\affiliation{Ioffe Institute, Politekhnicheskaya 26, St. Petersburg 194021, Russia}

\author{Christoffer Fremling}
\affiliation{Cahill Center for Astrophysics, 
California Institute of Technology, MC 249-17, 
1200 E California Boulevard, Pasadena, CA, 91125, USA}

\author[0000-0001-8205-2506]{V. Zach Golkhou}
\affiliation{DIRAC Institute, Department of Astronomy, University of Washington, 3910 15th Avenue NE, Seattle, WA 98195, USA} 
\affiliation{The eScience Institute, University of Washington, Seattle, WA 98195, USA}
\altaffiliation{Moore-Sloan, WRF Innovation in Data Science, and DIRAC Fellow}

\author{Matthew J. Graham}
\affiliation{Cahill Center for Astrophysics, 
California Institute of Technology, MC 249-17, 
1200 E California Boulevard, Pasadena, CA, 91125, USA}

\author{David Hale}
\affiliation{Caltech Optical Observatories, California Institute of Technology, Pasadena, CA  91125}

\author{George Helou}
\affiliation{
IPAC, California Institute of Technology, 1200 E. California Blvd, Pasadena, CA 91125, USA}

\author{Assaf Horesh}
\affiliation{
Racah Institute of Physics, The Hebrew University of Jerusalem, Jerusalem 91904, Israel}

\author{Mansi M.~Kasliwal}
\affiliation{Cahill Center for Astrophysics, 
California Institute of Technology, MC 249-17, 
1200 E California Boulevard, Pasadena, CA, 91125, USA}

\author[0000-0003-2451-5482]{Russ R. Laher}
\affiliation{
IPAC, California Institute of Technology, 1200 E. California Blvd, Pasadena, CA 91125, USA}

\author[0000-0002-8532-9395]{Frank J. Masci}
\affiliation{
IPAC, California Institute of Technology, 1200 E. California Blvd, Pasadena, CA 91125, USA}

\author[0000-0001-9515-478X]{A.~A.~Miller}
\affiliation{Center for Interdisciplinary Exploration and Research in Astrophysics and Department of Physics and Astronomy, Northwestern University, 1800 Sherman Ave, Evanston, IL 60201, USA}
\affiliation{The Adler Planetarium, Chicago, IL 60605, USA}

\author{Michael Porter}
\affiliation{Caltech Optical Observatories, California Institute of Technology, Pasadena, CA  91125}

\author{Anna Ridnaia}
\affiliation{Ioffe Institute, Politekhnicheskaya 26, St. Petersburg 194021, Russia}

\author[0000-0001-7648-4142]{Ben Rusholme}
\affiliation{
IPAC, California Institute of Technology, 1200 E. California Blvd, Pasadena, CA 91125, USA}

\author[0000-0003-4401-0430]{David L. Shupe}
\affiliation{
IPAC, California Institute of Technology, 1200 E. California Blvd, Pasadena, CA 91125, USA}

\author[0000-0001-6753-1488]{Maayane T. Soumagnac}
\affiliation{Lawrence Berkeley National Laboratory, 1 Cyclotron Road, Berkeley, CA 94720, USA}
\affiliation{Department of Particle Physics and Astrophysics, Weizmann Institute of Science, Rehovot 76100, Israel}

\author[0000-0002-2208-2196]{Dmitry S. Svinkin}
\affiliation{Ioffe Institute, Politekhnicheskaya 26, St. Petersburg 194021, Russia}

\correspondingauthor{Anna Y. Q. Ho}
\email{ah@astro.caltech.edu}

\submitjournal{The Astrophysical Journal}
\received{2 March 2020}
\revised{13 April 2020}

\begin{abstract}

We present \name\ (the ``Koala''), a fast blue optical transient discovered in the Zwicky Transient Facility (ZTF) One-Day Cadence (1DC) Survey. ZTF18abvkwla has a number of features in common with the groundbreaking transient AT\,2018cow: blue colors at peak ($g-r\approx-0.5$\,mag), a short rise time from half-max of under two days, a decay time to half-max of only three days, a high optical luminosity 
($M_{g,\mathrm{peak}}\approx-20.6$\,mag), a hot ($\gtrsim 40,000\,$K) featureless spectrum at peak light,
and a luminous radio counterpart.
At late times ($\Delta t>80\,\days$) the radio luminosity of ZTF18abvkwla ($\nu L_\nu \gtrsim 10^{40}\,\erg\,\psec$ at 10\,\ghz, observer-frame) is most similar to that of long-duration gamma-ray bursts (GRBs).
The host galaxy is a dwarf starburst galaxy ($M\approx5\times10^{8}\,\msol$, $\mathrm{SFR}\approx7\,\msol\,\pyr$) that is moderately metal-enriched ($\log\mathrm{[O/H]} \approx 8.5$),
similar to the hosts of GRBs and superluminous supernovae.
As in AT2018cow, the radio and optical emission in \name\ likely arise from two separate components:
the radio from fast-moving ejecta ($\Gamma \beta c >0.38c$) and the optical
from shock-interaction with confined dense material ($<0.07\,M_\odot$ in $\sim 10^{15}\,\cm$).
Compiling transients in the literature with
$\trise <5\,\days$ and $M_\mathrm{peak}<-20\,$mag,
we find that a significant number are engine-powered,
and suggest that the high peak optical luminosity is directly related to the presence of this engine.
From 18 months of the 1DC survey, we find that transients in this rise-luminosity phase space are at least two to three orders of magnitude less common than CC\,SNe. Finally, we discuss strategies for identifying such events with future facilities like the Large Synoptic Survey Telescope,
and prospects for detecting accompanying X-ray and radio emission.

\end{abstract}

\section{Introduction}
\label{sec:introduction}

Historically, the cadence of optical time-domain surveys
was tuned to detecting Type Ia supernovae (SNe),
whose optical light curves rise from first light to peak in 15--20\,days\ \citep{Miller2020}.
Recognizing that this observing strategy resulted in ``gaps'' in timescale-luminosity phase-space,
surveys such as the Palomar Transient Factory \citep{Law2009,Rau2009} and the Pan-STARRS1 Medium Deep Survey \citep{Drout2014}
sought to systematically chart the landscape of short-timescale ($<10\,$day) phenomena.
These efforts delineated
populations of fast transients spanning many orders of magnitude in peak luminosity,
from faint calcium-rich transients
\citep{Kasliwal2012} to luminous relativistic explosions
\citep{Cenko2013}.

A population of particular recent interest is ``fast evolving luminous transients'' \citep{Rest2018} or ``fast blue optical transients'' \citep{Margutti2019}.
A consistent definition of this ``class'' does not yet exist;
these terms typically refer to transients with rise times and peak luminosities too fast and too luminous, respectively, to be explained
by the radioactive decay of \nickel.
Although they likely arise from a variety of progenitors,
fast-luminous transients
are primarily found in star-forming galaxies \citep{Drout2014,Pursiainen2018} and therefore are thought to represent a variety of poorly understood endpoints of massive-star evolution.
As summarized in \citet{Kasen2017},
fast and luminous light curves
may be powered by
shock breakout or shock-cooling emission from material that is closely confined to the progenitor star at the time of explosion,
or alternatively by a ``central engine:''
accretion onto a black hole,
or the rotational spindown of a magnetar.

Most fast-luminous optical transients have been found in archival searches of optical-survey data, including PS1 \citep{Drout2014}, the Dark Energy Survey \citep{Pursiainen2018}, Kepler \citep{Rest2018}, and the Supernova Legacy Survey \citep{Arcavi2016}.
A handful have been discovered while the transient was still active, enabling prompt follow-up observations.
For example, spectroscopic monitoring of the fast-luminous transients iPTF16asu and ZTF18abukavn (SN\,2018gep) revealed that as the optical emission faded, the spectrum developed features typical of broad-lined Ic SNe \citep{Whitesides2017,Wang2019,Ho2019b}.

The discovery of the fast-luminous transient AT2018cow \citep{Prentice2018} generated considerable excitement because of its proximity ($z=0.0141$) and therefore the opportunity for detailed observations.
AT2018cow had several remarkable features:
(1) near-relativistic ejecta velocities at early times, from optical spectroscopy \citep{Perley2019cow}; (2) luminous and fast-varying X-ray emission suggesting an exposed central engine \citep{Rivera2018,Ho2019a,Margutti2019}; (3) high-velocity emission lines of hydrogen and helium emerging at late times \citep{Perley2019cow}; (4) no second peak that would indicate a significant role for radioactive ejecta in powering the light curve \citep{Perley2019cow}; and (5) luminous submillimeter emission indicating a large explosion energy injected into a shell of very dense material \citep{Ho2019a,Huang2019}.
Despite extensive observations across the electromagnetic spectrum,
the progenitor of AT2018cow is unknown.
One suggestion is a massive-star explosion that resulted in the formation of an accreting black hole or magnetar, which drove a mildly relativistic jet or wind \citep{Perley2019cow,Margutti2019,Ho2019a}. Other suggestions include an electron-capture SN \citep{Lyutikov2019} and a tidal disruption event (TDE; \citealt{Vinko2015,Perley2019cow,Kuin2019}).
If AT2018cow was a massive-star explosion,
the dense confined CSM points to eruptive mass-loss shortly before core-collapse \citep{Ho2019a},
and indeed \citet{Fox2019} pointed out the similarity between AT2018cow and interaction-powered Type Ibn SNe.

\begin{figure}[htb!]
    \centering
    \includegraphics[width=1.0\columnwidth]{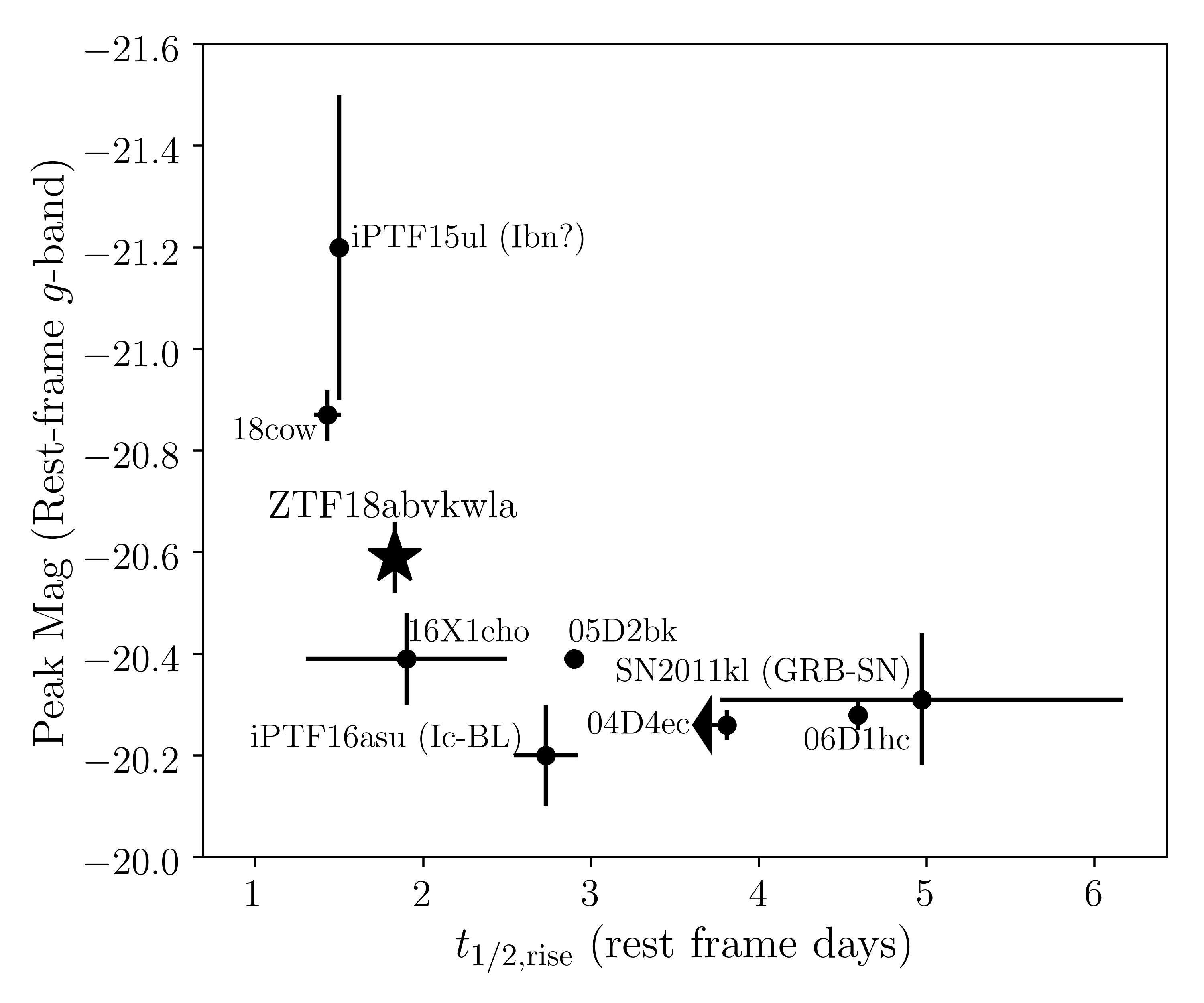}
    \caption{Phase-space of luminosity and rise time considered in this paper;
    see Table~\ref{tab:literature} for data sources.
    We do not show the transient Dougie \citep{Vinko2015}, which had a peak absolute magnitude of $-23$.
    Note that the peak mag of iPTF15ul includes a large host-galaxy extinction correction, whereas the other sources have zero host extinction correction.
    Also note that SN\,2011kl was associated with an ultra-long duration GRB\,111209A \citep{Kann2018}, and the light-curve properties shown here reflect the afterglow-subtracted light curve \citep{Kann2019}.}
    \label{fig:lum-rise}
\end{figure}

Here we report the discovery in Zwicky Transient Facility (ZTF) data of \name\footnote{nicknamed ``Koala'' on account of the last four letters of its ZTF ID}, a fast-rising luminous optical transient at $z=0.27$.\footnote{After the submission of our paper,
\citet{Coppejans2020} published radio and X-ray observations of CSS161010,
another transient in a dwarf galaxy with properties similar to AT2018cow.}
In \S \ref{sec:observations}
we present the key observational features of \name---a rest-frame $g$-band light curve similar to that of AT2018cow,
a luminous radio counterpart similar to gamma-ray burst (GRB) afterglows, and
a starburst dwarf host galaxy.
In \S \ref{sec:comparison} we compare ZTF18abvkwla to transients in the literature
that have $\trise<5\,\days$ and $M<-20$\,mag, where \trise\ is defined from 0.75\,mag below peak to peak (half-max to max in flux space).
We use a cut of $M<-20\,$mag to exclude ``normal'' Type Ibn SNe \citep{Hosseinzadeh2017} and
we exclude the hundreds of optical afterglows discovered in GRB follow-up observations \citep{Kann2010}.
The comparison sample is shown in Table~\ref{tab:literature} and Figure~\ref{fig:lum-rise}.
Note that the Table~\ref{tab:literature} transients have thermal spectra at peak,
unlike GRB afterglows which arise from synchrotron radiation.

\begin{deluxetable*}{lrrrrr}[htb!]
\tablecaption{Transients in the literature with $\trise<5\,\days$ and $M<-20\,$mag. Timescales are presented in rest-frame and measured using the light curve that most closely matches rest-frame $g$.
Luminosity is corrected for Galactic extinction, assuming zero host-galaxy extinction in all cases except for iPTF15ul and SN\,2011kl.
SN\,2011kl was associated with GRB\,111209A, and the afterglow emission has been subtracted.
\label{tab:literature}
}
\tabletypesize{\scriptsize}
\startdata
\tablehead{
Name & Redshift & $M_{g,\mathrm{max}}$ & \trise & \tfade & Ref \\
 & & & days & days}
Dougie & 0.19 & $-23.03\pm0.13$ & $3.92\pm0.14$ & $9.69\pm1.19$ & [1] \\
SN\,2011kl & 0.677 & $-20.31\pm0.13$ & $4.97\pm1.20$ & $17.70\pm5.82$ & [2,3] \\
SNLS04D4ec & 0.593 & $-20.26\pm0.03$ & $<3.81$ & $8.60\pm0.43$ & [4] \\
SNLS05D2bk & 0.699 & $-20.39\pm0.02$ & $2.90\pm0.06$ & $12.75\pm0.78$ & [4] \\
SNLS06D1hc & 0.555 & $-20.28\pm0.03$ & $4.59\pm0.06$ & $12.35\pm0.45$ & [4] \\
iPTF15ul & 0.066 & $-21.2 \pm 0.3$ & $1.53\pm0.05$ & $3.72\pm0.08$ & [5] \\
DES16X1eho & 0.76 & $-20.39 \pm 0.09$ & 1.28--2.53 & $1.01 \pm 0.27$ & [6] \\
iPTF16asu & 0.187 & $-20.3\pm0.1$ & $1.14\pm0.13$ & $10.62\pm0.55$ & [7] \\
AT2018cow & 0.0141 & $-20.89\pm0.06$ & $1.43\pm0.08$ & $1.95\pm0.06$ & [8,9] \\
\enddata
\tablereferences{
[1] \citet{Vinko2015}, [2] \citet{Greiner2015}, [3] \citet{Kann2019}, [4] \citet{Arcavi2016}, [5] \citet{Hosseinzadeh2017}, [6] \citet{Pursiainen2018} [7] \citet{Whitesides2017}, [8] \citet{Prentice2018} [9] \citet{Perley2019cow}
}
\end{deluxetable*}

In \S \ref{sec:lc-modeling}
we model the optical emission from \name\ as thermal emission from shock breakout in dense confined material,
and in \S \ref{sec:modeling-radio} we use the radio emission to estimate properties of the forward shock (velocity, shock energy) and the ambient medium.
In \S \ref{sec:progenitors},
we discuss possible progenitor systems.
Finally, in \S \ref{sec:rates} we use 18 months of survey observations to estimate the rate of transients in the phase-space of Figure~\ref{fig:lum-rise},
and find that the rate is 2--3 times smaller than the CC SN rate.

Throughout this paper, we use a standard $\Lambda$CDM cosmology \citep{Planck2016} and times are reported in UT.
Optical magnitudes are reported in the AB system \citep{Oke1983}, and corrected for foreground Galactic extinction using reddening measurements in \citet{Schlafly2011} and the extinction law from \citet{Fitzpatrick1999}.

\section{Discovery and Basic Analysis}
\label{sec:observations}

\subsection{Optical}
\label{sec:obs-optical}

\subsubsection{Photometry}
\label{sec:obs-opt-phot}

\begin{figure}[htb!]
    \centering
    \includegraphics[width=\columnwidth]{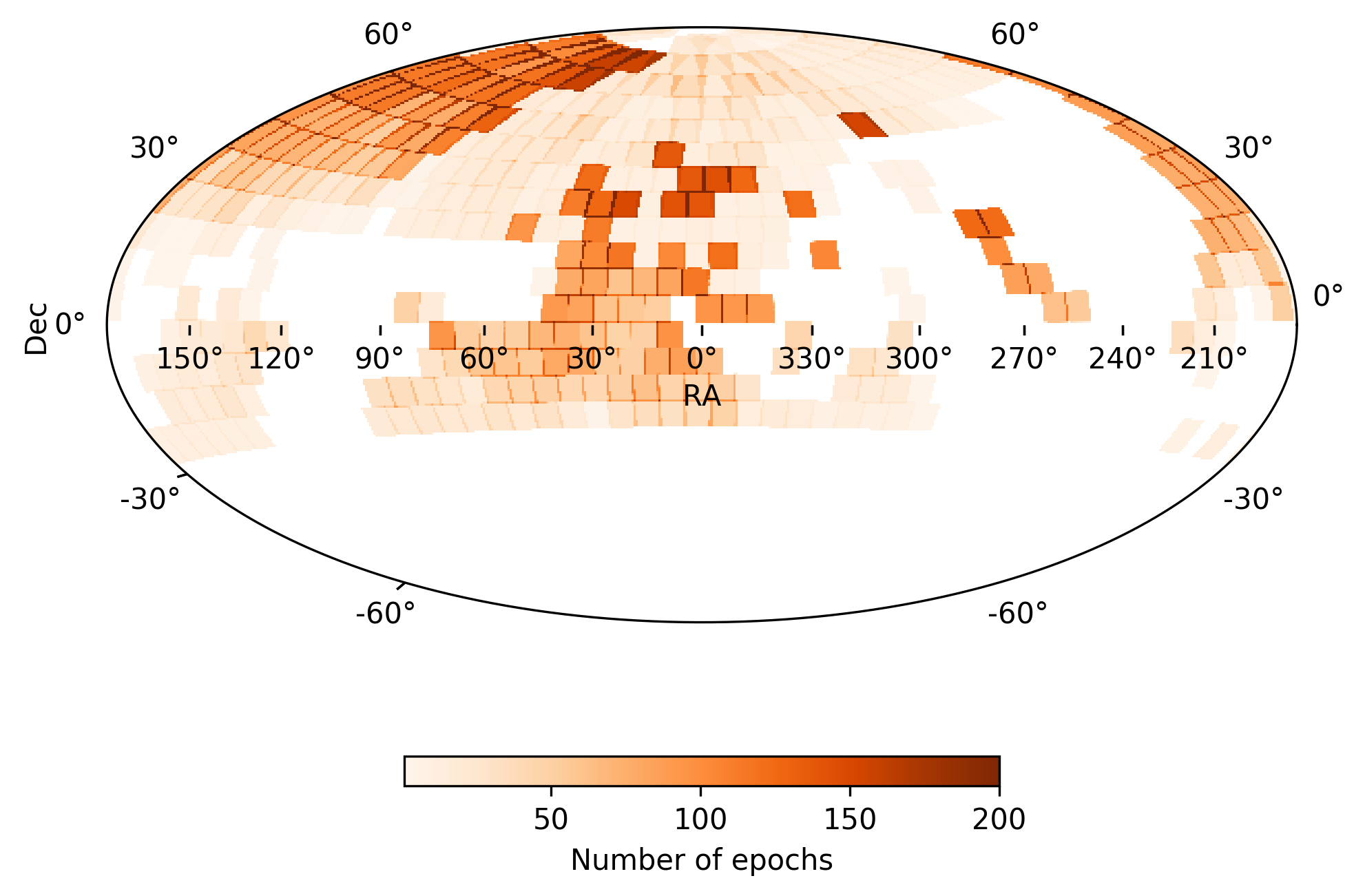}
    \includegraphics[width=\columnwidth]{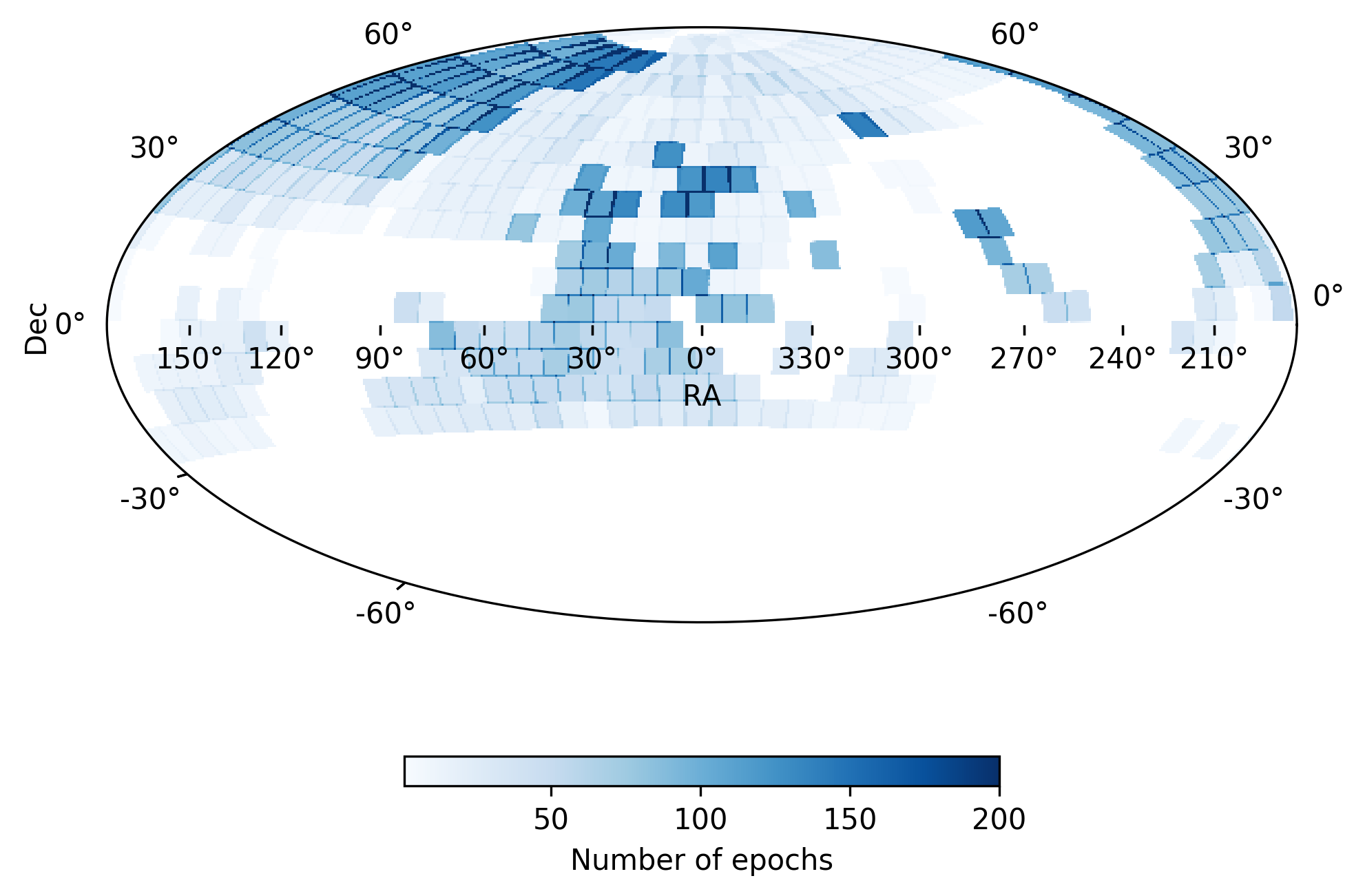}
    \caption{Number of epochs obtained by the ZTF one-day cadence survey from 3 April 2018 to 18 October 2019}
    \label{fig:skycoverage}
\end{figure}

\begin{figure}[htb!]
    \centering
    \includegraphics[width=1.0\columnwidth]{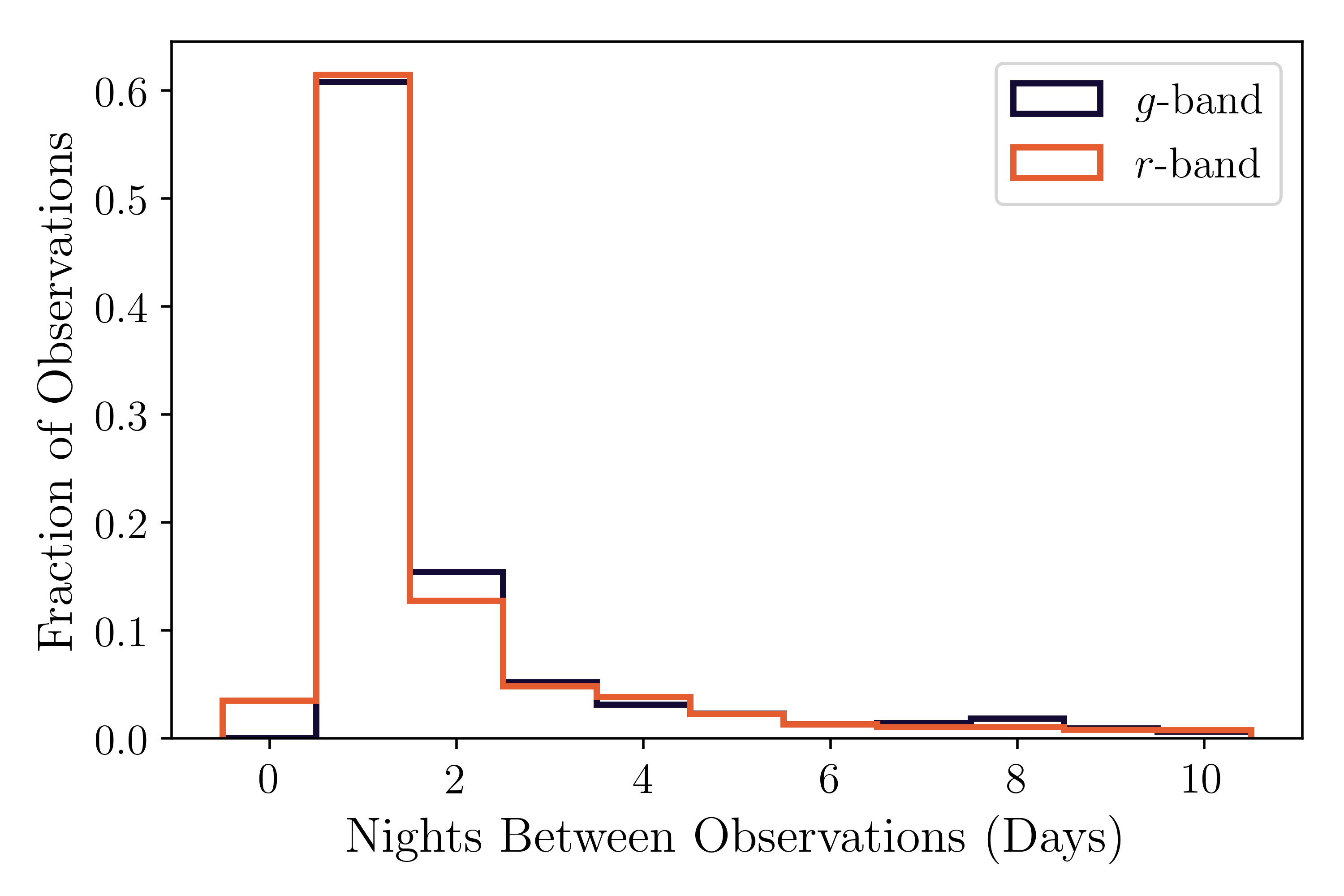}
    \caption{Histogram of times between successive observations of a field in the same filter for the ZTF one-day cadence survey. Intervals greater than 10 days are not shown.}
    \label{fig:cadence-hist}
\end{figure}

Since April 2018,
ZTF \citep{Bellm2019a,Graham2019}
has been conducting a wide-area (2000--3000\,\degsq) one-day cadence (1DC) survey in $g$ and $r$ \citep{Bellm2019b}.
The sky coverage of the 1DC survey is shown in Figure~\ref{fig:skycoverage}
and a histogram of the typical time between exposures is shown in Figure~\ref{fig:cadence-hist}.

The IPAC ZTF pipeline \citep{Masci2019} uses
the method described in \citet{Zackay2016}
to generate difference images using a coadded reference image.
Every 5$\sigma$ point-source detection is
assigned a score based on a machine learning real-bogus metric \citep{Mahabal2019,Duev2019},
and is cross-matched against external catalogs to search for resolved and extended counterparts \citep{Tachibana2018}.
Alerts are distributed in Avro format \citep{Patterson2019}
and are filtered by the ZTF collaboration
using a web-based system called the GROWTH Marshal \citep{Kasliwal2019}.

\name\ was discovered in an image obtained on 12 Sept 2018.
The alert passed a filter designed to look for rapidly-evolving transients,
and as a result we obtained a follow-up spectrum 24 hours later (\S \ref{sec:obs-opt-spec}).
The discovery magnitude was $g=19.73 \pm 0.16$ mag
and the last non-detection was one day prior, with a limiting magnitude $g>20.74$.

The source position was measured to be
$\alpha = 02^{\mathrm{h}}00^{\mathrm{m}}15.19^{\mathrm{s}}$, $\delta = +16^{\mathrm{d}}47^{\mathrm{m}}57.3^{\mathrm{s}}$ (J2000), which is
$0\farcs28\pm0\farcs13$ from the nucleus of a blue ($g-r=0.32\,$mag) extended source that has a photometric redshift of 0.11 (68 percentile confidence interval 0.08--0.29) in the eighth data release of LegacySurvey (DR8; \citealt{Dey2019}).
At $z=0.2714$ (\S \ref{sec:obs-opt-spec})
this offset corresponds to $1.9\pm0.9\,\kpc$.
The host is approximately $2''$ (14\,\kpc) across.

\begin{figure*}[htb!]
\includegraphics[width=\textwidth]{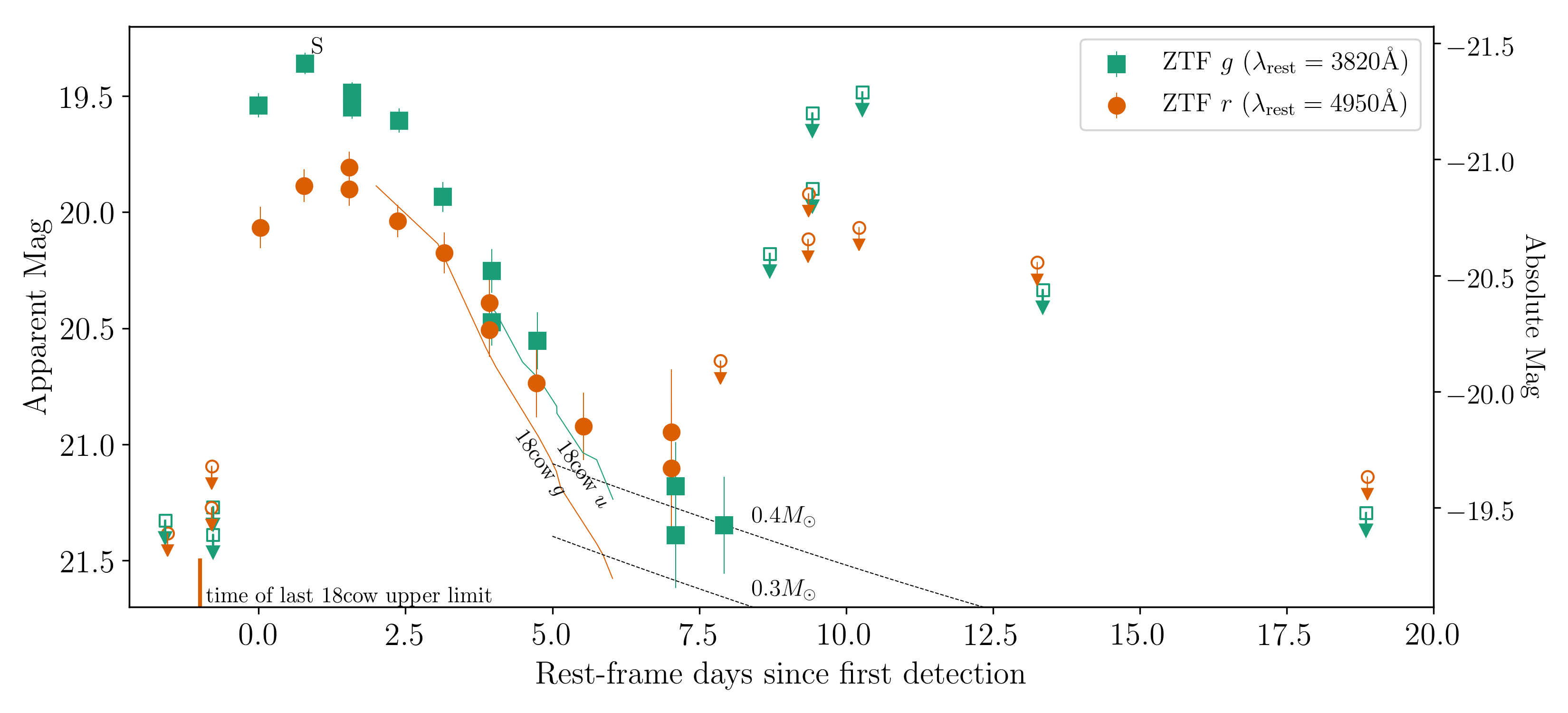}
\caption{Light curve of \name\ in P48 $g$ (filled green squares) and $r$ (open orange circles) with a comparison to AT2018cow at similar rest wavelengths,
both corrected for Galactic extinction.
The `S' at the top of the inset indicates the epoch of our DBSP spectrum.
Dashed lines show \nickel-powered light curves for two different nickel masses.
\label{fig:lightcurve}}
\end{figure*}

The light curve (Figure~\ref{fig:lightcurve}; Table~\ref{tab:opt-phot}) has a similar timescale and peak luminosity to that of AT2018cow.
In rest-frame $g$-band,
the rise time is $1.83\pm0.05\,\days$,
the fade time is $3.12\pm0.22\,\days$,
and the peak magnitude is $-20.59\pm0.07$\,mag.

\begin{deluxetable}{lrrr}[htb!]
\tablecaption{Optical photometry for \name\ from forced photometry on P48 images \citep{Yao2019}. Values have not been corrected for Galactic extinction. Phase $\Delta t$ is defined from $t_0$, the last non-detection. \label{tab:opt-phot}} 
\tablewidth{0pt} 
\tablehead{ \colhead{Date (MJD)} & \colhead{$\Delta t$} & \colhead{Filter} & \colhead{AB Mag} } 
\tabletypesize{\scriptsize} 
\startdata 
58372.39 & $-1.02$ & $r$ & $<21.39$ \\
58372.42 & $-0.99$ & $g$ & $<21.56$ \\
58373.41 & 0.00 & $g$ & $19.71 \pm 0.05$ \\ 
58373.45 & 0.04 & $r$ & $20.18 \pm 0.09$ \\ 
58374.39 & 0.98 & $r$ & $20.00 \pm 0.07$ \\ 
58374.41 & 1.00 & $g$ & $19.53 \pm 0.05$ \\ 
58375.37 & 1.96 & $r$ & $19.92 \pm 0.07$ \\ 
58375.37 & 1.96 & $r$ & $20.02 \pm 0.07$ \\ 
58375.43 & 2.03 & $g$ & $19.65 \pm 0.04$ \\ 
58375.43 & 2.03 & $g$ & $19.72 \pm 0.05$ \\ 
58376.42 & 3.01 & $r$ & $20.15 \pm 0.07$ \\ 
58376.44 & 3.04 & $g$ & $19.77 \pm 0.05$ \\ 
58377.39 & 3.98 & $g$ & $20.10 \pm 0.07$ \\ 
58377.43 & 4.02 & $r$ & $20.29 \pm 0.09$ \\ 
58378.40 & 4.99 & $r$ & $20.50 \pm 0.10$ \\ 
58378.40 & 4.99 & $r$ & $20.62 \pm 0.12$ \\ 
58378.45 & 5.04 & $g$ & $20.64 \pm 0.10$ \\ 
58378.45 & 5.05 & $g$ & $20.42 \pm 0.09$ \\ 
58379.42 & 6.02 & $r$ & $20.85 \pm 0.15$ \\ 
58379.44 & 6.04 & $g$ & $20.72 \pm 0.12$ \\ 
58380.43 & 7.03 & $r$ & $21.04 \pm 0.15$ \\ 
58382.34 & 8.93 & $r$ & $21.06 \pm 0.27$ \\ 
58382.34 & 8.93 & $r$ & $21.22 \pm 0.28$ \\ 
58382.43 & 9.03 & $g$ & $21.35 \pm 0.19$ \\ 
58382.43 & 9.03 & $g$ & $21.56 \pm 0.23$ \\ 
58383.48 & 10.07 & $g$ & $21.51 \pm 0.21$ \\ 
\enddata 
\end{deluxetable}

We estimate that the onset of the optical emission was around the time of the last non-detection ($t_0=2458372.9206$ JD) and use this as a reference epoch for the remainder of the paper.

\subsubsection{Spectroscopy and Host Galaxy Properties}
\label{sec:obs-opt-spec}

\begin{figure*}[htb!]
\includegraphics[width=\textwidth]{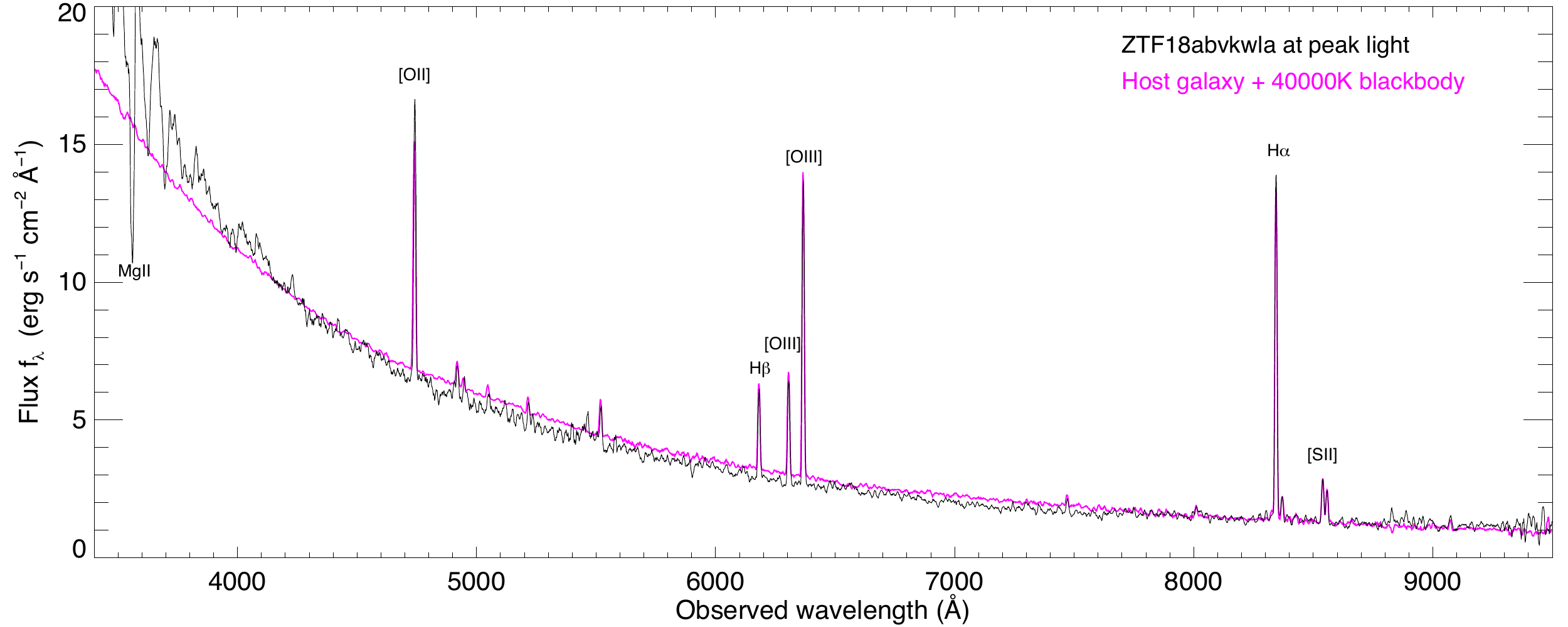}
\caption{The spectrum of \name\ at the peak of the $g$-band optical light curve (black), which was 1 day after the first detection. The source is extremely hot and blue with no spectral features except those associated with the host galaxy. Overplotted in pink is a rescaled late-time spectrum of the host galaxy with a $40,000\,$\,K blackbody added.}
\label{fig:dbspspec}
\end{figure*}

One day after discovery,
we obtained a spectrum of \name\
using the Double Beam Spectrograph (DBSP; \citealt{Oke1982}) on the 200-inch Hale telescope at Palomar Observatory.
We used the D55 dichroic,
a slit width of 1.5 arcseconds,
the 600/4000 blue grating,
and the 316/7500 red grating.
The spectrum was reduced using a PyRAF-based pipeline
\citep{Bellm2016}.
As  shown in Figure~\ref{fig:dbspspec},
the spectrum shows a hot blue continuum with no broad features in emission or absorption.  Superimposed on the spectrum are a variety of narrow emission lines typical of a star-forming galaxy (H$\alpha$, H$\beta$, \ion{O}{3}, \ion{S}{2}, \ion{O}{2}) at a redshift of $z=0.2714$ plus the \ion{Mg}{2} UV doublet in absorption at consistent redshift.

A blackbody fit to the continuum (after subtracting a host-galaxy continuum model, discussed later in this section) indicates an effective temperature $T \gtrsim 40,000$K,
although we caution that it could be significantly higher as
the bulk of the energy was clearly emitted in the UV ($<2750\,$\AA\ in the rest frame) and we have no firm constraint on the host-galaxy extinction.
Together with the peak absolute magnitude of the $g$-band light curve,
we derive a bolometric luminosity of $ L_\mathrm{bol}> \nu L_\nu \sim 10^{44}\,\erg\,\psec$.
Assuming $T=40,000\,$K, the photospheric radius is
$R>2 \times 10^{14}\,\cm$.
Since the peak is 2\,\days\ after first light,
assuming $R(t=t_0)=0$ gives $V>0.04c$.

On 4 Jan 2019 (+115\,\days), we obtained a spectrum of the host galaxy of \name\ using the Low Resolution Imaging Spectrometer \citep{Oke1995,McCarthy1998} on the Keck I 10-m telescope,
with the 400/3400 grism in the blue camera and the 400/8500 grating in the red camera. Exposure times were 940 and 900 seconds for the blue and red camera respectively. The spectrum was reduced and extracted using \texttt{Lpipe} \citep{Perley2019lpipe}.
The absolute calibration was established independently for each camera (red vs. blue) by calculating synthetic photometry of the output spectra in the blue and red cameras in the $g$ and $r$ bands, respectively, and rescaling to match the $g$ and $r$ photometry from SDSS DR14 \citep{Abolfathi2018}.
The SDSS magnitudes (AB, converted to Pogson) are $u = 21.74 \pm 0.20$, $g = 21.20 \pm 0.04$, $r = 20.81 \pm 0.05$, $i = 20.92 \pm 0.09$, and $z = 20.52 \pm 0.20$.

\begin{figure*}[htb!]
\includegraphics[width=\textwidth]{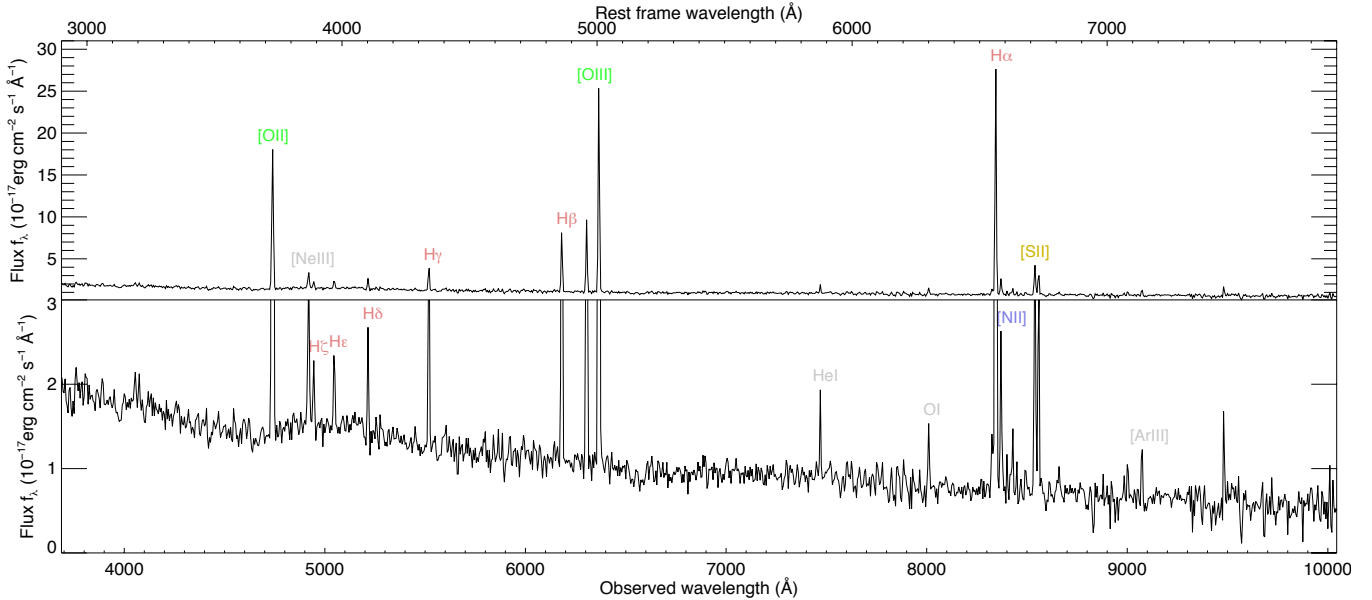}
\caption{
Spectrum of the host galaxy of \name.  The scale on the bottom half has been zoomed in to show the galaxy continuum and weak emission lines.
The feature at 9500\,\AA\ is a sky-subtraction residual.
}
\label{fig:hostspec}
\end{figure*}

The host-galaxy spectrum (Figure~\ref{fig:hostspec}) consists of a weak continuum and a series of very strong emission lines.  Line fluxes were extracted using an identical procedure as in \cite{Perley2016}.  We first fit a model to the spectral energy distribution (SED). We used a custom IDL routine based on the templates of \citet{Bruzual2003} to fit the SDSS $ugriz$ photometry, including the contribution of nebular lines.
As only SDSS $ugriz$ photometry is available to fit the host-galaxy SED it is difficult to constrain the nature of the stellar population of the host galaxy in detail, and we were only able to fit the simplest possible model (a continuous star-formation history).  However, the stellar mass is unambiguously low ($\sim 5 \times 10^8\,\msol$, comparable to the SMC),

This model was then used to produce a synthetic galaxy continuum spectrum, which was subtracted from the observed one (this correction is significant only for higher-order Balmer lines, which overlay strong galaxy absorption features).  Emission line fluxes were then measured by fitting a Gaussian function to each emission line (plus a linear baseline to fit any continuum residuals). Lines that were blended or very nearby were fit in groups, and lines whose ratios are fixed from theory were tied together in fitting.
A list of all measured line fluxes is given in Table~\ref{tab:lineflux}.  

\begin{deluxetable}{lrrr}[htb!]
\tablecaption{
    Host emission line fluxes and equivalent widths
    \label{tab:lineflux}
}
\tablehead{
\colhead{Species} & \colhead{Wavelength} & \colhead{Flux} & \colhead{Eq. Width}
\\
 & \colhead{(\AA)} & \colhead{(erg cm$^{-2}$s$^{-1})$} & \colhead{(\AA)}
}
\startdata
H$\alpha$ & 6562.82 &214.74 $\pm$ 2.71 &  205.9 $\pm$ 7.0 \\
H$\beta$  & 4861.33 & 57.57 $\pm$ 1.07 & 41.3 $\pm$ 1.1 \\
H$\gamma$ & 4340.47 & 26.98 $\pm$ 1.03 & 17.6 $\pm$ 0.8 \\
H$\delta$ & 4101.74 & 13.92 $\pm$ 0.91 &  7.2 $\pm$ 0.5 \\
H$\epsilon$&3970.08 & 11.44 $\pm$ 0.86 &  5.9 $\pm$ 0.4 \\
H$\zeta$ & 3889.06 &   9.72 $\pm$ 0.88 &  5.0 $\pm$ 0.5 \\
 \OII    &  3727    & 159.44 $\pm$ 1.72 & 89.0 $\pm$ 2.4 \\ 
 \NeIII  & 3868.76 &  16.00 $\pm$ 0.94 &   8.3 $\pm$ 0.5 \\
 \OIII   & 4363.21  &  $<$3.31          &   $<$2.1        \\
 \OIII   & 4958.91  &  66.35 $\pm$ 1.37 &  47.6 $\pm$ 1.4 \\
 \OIII   & 5006.84  & 196.88 $\pm$ 1.60 & 141.3 $\pm$ 3.1 \\
 \HeI    & 5875.62 &   6.76 $\pm$ 0.72 &   5.7 $\pm$ 0.6  \\
 \NII    & 6548.06  &   4.90 $\pm$ 0.69 &   4.7 $\pm$ 0.7 \\
 \NII    & 6583.57  &  13.91 $\pm$ 0.82 &  13.3 $\pm$ 0.9 \\
 \SII    & 6716.44  &  27.86 $\pm$ 0.95 &  29.3 $\pm$ 1.2 \\
 \SII    & 6730.82  &  21.81 $\pm$ 0.71 &  22.9 $\pm$ 0.9 \\ 
 \OI     & 6300.30 &   6.76 $\pm$ 0.71 &   7.2 $\pm$ 0.8  \\
 \ArIII  & 7135.79 &   5.49 $\pm$ 0.53 &   6.8 $\pm$ 0.7 \\
\enddata
\end{deluxetable}


The SED fitting and the emission-line analysis produce consistent estimates of $7\,\msol\,\pyr$ for the star-formation rate,
and a very high specific 
star-formation rate of 
$\sim$10$^{-8}\,\pyr$.
This implies a stellar population dominated by young stars formed in a recent triggered star-formation burst episode.

We used the host galaxy spectrum
(Figure~\ref{fig:hostspec}) to calculate
standard emission-line diagnostics, including metallicity estimates on a variety of scales using the Monte-Carlo code of \citet{Bianco2016}.
These metallicity measurements are provided in Table~\ref{tab:metallicities}.
The basic properties of the host galaxy are listed in Table~\ref{tab:hostsedprop}.

\begin{deluxetable}{lr}[htb!]
\tablecaption{
    Host galaxy properties (metallicities, mainly) from PyMCZ.
    \label{tab:metallicities}
}
\tablehead{}
\startdata
 SFR$^\tablenotemark{a}$   &            $6.47\pm1.3$ \\
 E(B-V)         &            $0.220_{-0.022}^{+0.023}$ \\
 logR23         &            $0.903_{-0.012}^{+0.012}$ \\ 
 D02            &            $8.253_{-0.128}^{+0.130}$ \\
 Z94            &            $8.450_{-0.010}^{+0.016}$ \\
 M91            &            $8.219_{-0.026}^{+0.026}$ \\
 PP04\_N2Ha     &            $8.200_{-0.010}^{+0.010}$ \\
 PP04\_O3N2     &            $8.187_{-0.009}^{+0.008}$ \\
 P10\_ONS       &            $8.708_{-0.024}^{+0.024}$ \\
 P10\_ON        &            $8.172_{-0.047}^{+0.046}$ \\
 M08\_N2Ha      &            $8.361_{-0.021}^{+0.020}$ \\
 M08\_O3O2      &            $8.521_{-0.011}^{+0.011}$ \\
 M13\_O3N2      &            $8.174_{-0.009}^{+0.009}$ \\
 M13\_N2        &            $8.194_{-0.042}^{+0.041}$ \\
 KD02\_N2O2     &            $7.567_{-0.074}^{+0.722}$ \\ 
 KK04\_N2Ha     &            $8.381_{-0.029}^{+0.028}$ \\
 KK04\_R23      &            $8.390_{-0.021}^{+0.021}$ \\ 
 KD02comb       &            $8.304_{-0.024}^{+0.024}$ \\
\enddata
\tablenotetext{a}{SFR is not from PyMCZ but is calculated directly from the corrected Balmer-line fluxes based on the relation of \cite{Kennicutt+1994}.}
\end{deluxetable}

\begin{table}[htb!]
    \centering
    \caption{Properties of the host galaxy of \name. The stellar mass, star-formation rate, maximum age, and extinction are from a fit to the galaxy SED; the $\chi^2$ refers to that fit. The metallicity [O/H] was measured using the host galaxy spectrum and is provided on the Z94 scale. This value corresponds to $0.6\times$ solar.}
    \label{tab:hostsedprop}
\begin{tabular}{| l | l | l |}
\hline
Stellar mass        & M &   5.1$^{+3.4}_{-2.0}$ $\times$ $10^8$\,\msol\\
Star-formation rate &  SFR &   6.8$^{+3.7}_{-4.6}$\,\msol\,\pyr\\
Maximum age         &  age &   7.5$^{+30}_{-4.5}$ $\times 10^7$ yr \\
Extinction          &   Av &   0.72$^{+0.17}_{-0.54}$ mag  \\
                    & $\chi^2$/dof &       1.6 / 2 \\
Metallicity & 12+log[O/H] & 8.5 \\                    
\hline
\end{tabular}
\tablenotetext{}{\begin{flushleft}The SFR listed here is derived from the photometry, while the SFR in Table~\ref{tab:metallicities} was derived from the spectrum. So, there is no expectation of identical values or errors.\end{flushleft}}
\end{table}

\subsection{Radio Observations}
\label{sec:radio-obs}

We obtained four epochs of observations of \name\ using the Karl G. Jansky Very Large Array (VLA; \citealt{Perley2011}) 
under the program VLA/18B-242 (PI: D. Perley), listed in Table~\ref{tab:radio-log}.
The first epoch was at $\Delta t\approx81\,\days$ at X-band,
while the VLA was in C configuration.
We used 3C138 as our flux density and bandpass calibrator, and J0204+1514 as our complex gain calibrator.
The next three epochs were at $\Delta t\approx310\,\days$, $\Delta t\approx350\,\days$, and $\Delta t\approx400\,\days$,
all while the VLA was in A configuration.
We continued to use 3C138 but switched to J0238+1636 as our complex gain calibrator.
For each observation, we ran the standard VLA calibration pipeline
available in the Common Astronomy Software Applications (CASA; \citealt{McMullin2007}).
After calibration, we inspected the data manually for further flagging.
We imaged the data using the CLEAN algorithm \citep{Hogbom1974} available in CASA,
using a cell size that was 1/5 of the synthesized beamwidth.
The field size was set to be the smallest magic number ($10 \times 2^n$) larger than the number of cells needed to cover the primary beam.

\begin{deluxetable}{lrrrrr}[htb!]
\tablecaption{
    Radio observations of \name\ with the VLA and the GMRT. Upper limit is reported as $3\times$ the image RMS.} \label{tab:radio-log}
\tablehead{
\colhead{$\Delta t$} & \colhead{Facility} & \colhead{Obs. Date} & \colhead{Config.} & \colhead{$\nu$} & \colhead{Flux Density} \\
days &  & (UT) & & (GHz) & (mJy)
}
\startdata
81 & VLA & 2018-12-01 & C & 10 & $0.364 \pm 0.006$ \\
188 & VLA$^{a}$ & 2019-03-19 & B & 3 & $<0.134$ \\
310 & VLA & 2019-07-19 & BnA & 10 & $0.061 \pm 0.003$ \\
343 & VLA & 2019-08-21 & A & 6 & $0.089 \pm 0.003$ \\
346 & VLA & 2019-08-24 & A & 3 & $0.068 \pm 0.004$ \\
351 & VLA & 2019-08-29 & A & 1.5 & $0.146 \pm 0.013$ \\
352 & VLA & 2019-08-30 & A & 10 & $0.045 \pm 0.003$ \\
364 & GMRT & 2019-09-11 & -- & 0.6 & $<0.105$ \\
396 & VLA & 2019-10-13 & A & 10 & $0.031 \pm 0.003$ \\
397 & VLA & 2019-10-14 & A & 6 & $0.033 \pm 0.003$ \\
\enddata
\tablenotetext{a}{From VLASS}
\end{deluxetable}

In addition, the position of ZTF18abvkwla was serendipitously covered by
the VLA Sky Survey (VLASS; \citealt{Lacy2019}),
which has been mapping the entire sky visible to the VLA at low frequencies (2--4\,GHz)
in three epochs at a cadence of 32 months.
The Quicklook images are now available for the first epoch (17,000\,deg$^{-2}$).
We searched the existing Quicklook data using code available on Github\footnote{https://github.com/annayqho/Query\_VLASS} that locates the appropriate VLASS tile and subtile for a given RA and Dec
and extracts a cutout 12 arcsec on a side.
Given a non-detection we estimated an upper limit on the flux density by taking the standard deviation of the pixel values in this cutout, after performing initial 
3$\sigma$ clipping (removing pixels with a value greater than $3 \times$ the standard deviation).
The VLASS observation of ZTF18abvkwla is also listed in Table \ref{tab:radio-log}.

We obtained one epoch of observations with the upgraded Giant Metrewave Radio Telescope (GMRT; \citealt{Gupta2017,Swarup1991}) under a proposal for Director's Discretionary Time (Proposal \# ddtC086; PI: A. Ho).
For our GMRT observations,
we used 3C147 and 3C48 as our flux density and bandpass calibrators and 0238+166 for our phase calibrator.
We calibrated the GMRT data manually using commands in CASA,
with 6 rounds of phase-only self-calibration and 2 rounds of amplitude and phase self-calibration.

\begin{figure}[htb!]
    \centering
    \includegraphics[width=\columnwidth]{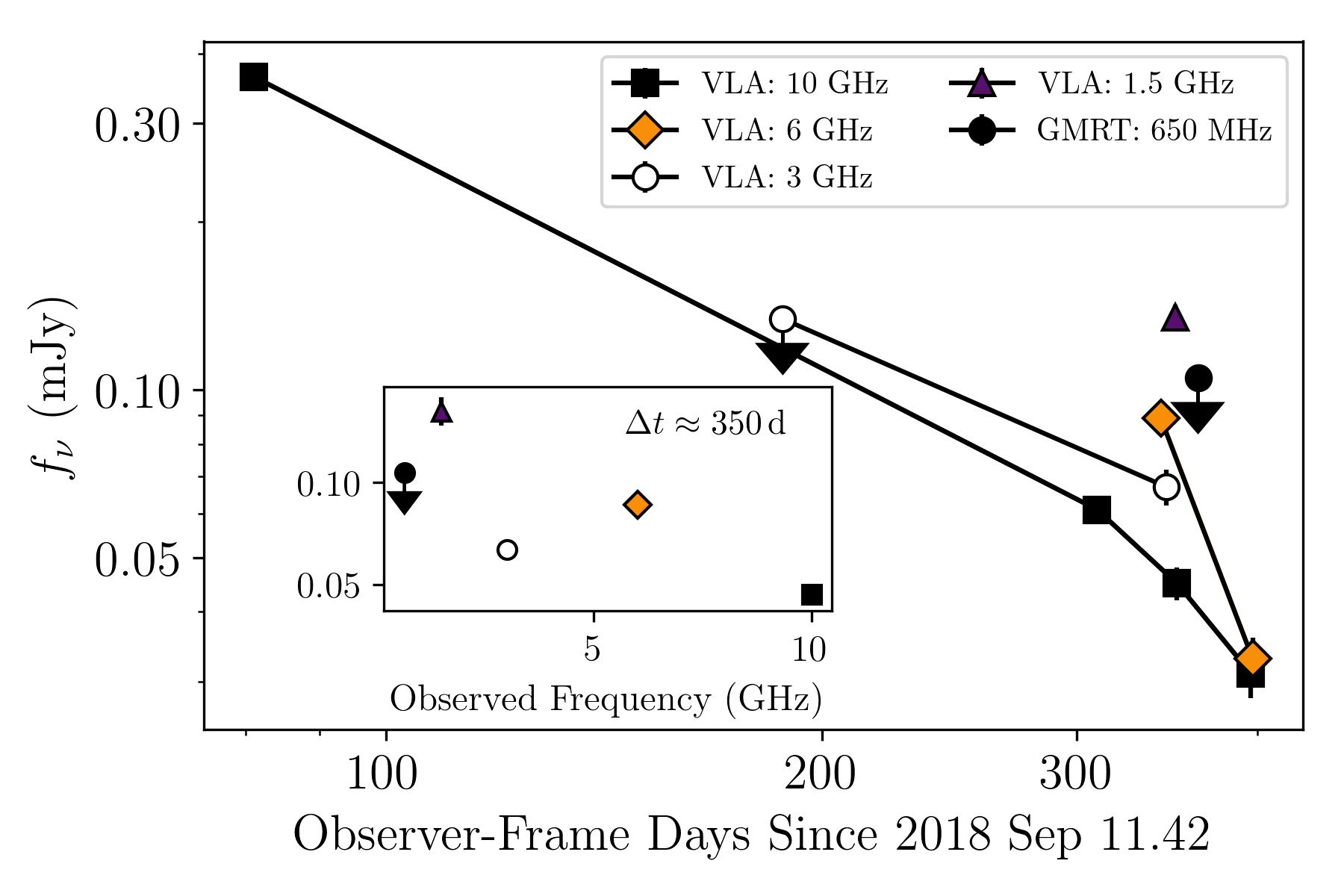}
    \caption{The radio light curve of \name\ with the spectral energy distribution at $\Delta t\approx350\,\days$ (rest-frame $\Delta t\approx275\,\days$) shown inset. The upper limit at 3\,\ghz\ comes from a serendipitous observation by VLASS.}
    \label{fig:radiolc}
\end{figure}

The radio light curve from the VLA is shown in Figure~\ref{fig:radiolc}.
At the time of our first observation,
the 10\,\ghz\ (rest-frame 12\,\ghz) luminosity was $10^{30}\,\erg\,\psec\,\phz$,
and the in-band spectral index was
$\alpha=-0.16 \pm 0.05$ where $F_\nu \propto \nu^{-\alpha}$.
At late times ($\Delta t>300\,\days$) the decline is very steep:
at 6\,\ghz\ we find $F_\nu \propto t^{-6.8\pm0.9}$,
and at 10\,\ghz\ we find $F_\nu \propto t^{-3.2\pm1.4}$.

To estimate the contribution to the radio emission from the host galaxy,
we use the relation in \citet{Greiner2016},
adapted from \citet{Murphy2011}:

\begin{equation}
\begin{split}
    \left( \frac{\mathrm{SFR}_\mathrm{Radio}}{\msol\,\pyr} \right) 
    & = 0.059
    \left( \frac{F_\nu}{\ujy} \right) (1+z)^{-(\alpha+1)} \\
    & \times \left( \frac{D_L}{\mathrm{Gpc}} \right)^2 \left( \frac{\nu}{\ghz} \right)^{-\alpha}.
\end{split}
\end{equation}

\noindent In the final epoch of our radio observations,
assuming $\alpha=-0.75$ \citep{Condon1992} where $F_\nu \propto \nu^{\alpha}$,
the 10\,\ghz\ flux density of $0.031\pm0.003\,$mJy
predicts a SFR of 20\,\msol\,\pyr.
So, we conclude that during the final observation the radio emission is still dominated by the transient,
but the host may contribute a non-trivial fraction of the flux.

\section{Comparison With Extragalactic Explosions}
\label{sec:comparison}

\subsection{Optical Light Curve and Spectrum}
\label{sec:comparison-optical}

As shown in \S \ref{sec:introduction},
the fast rise time and high peak luminosity of \name\ is shared by only a handful of transients in the literature.
In this section we compare the optical properties of \name\ to the transients in Table \ref{tab:literature}.
We exclude Dougie because it resided in an old stellar population with no signs of enhanced star formation \citep{Vinko2015};
the dominance of absorption features and much lower star-formation rate were confirmed by additional LRIS spectroscopy \citep{Arcavi2016}.

To compare light curves,
we selected the light curve in a filter closest to rest-frame $g$ (the same filters used in constructing Figure~\ref{fig:lum-rise}).
Following \citet{Whitesides2017},
we calculated absolute magnitudes using

\begin{equation}
   M = m_\mathrm{obs} - 5\,\log_{10} 
   \left( \frac{D_L}{10\,\mathrm{pc}} \right) 
   + 2.5\,\log_{10}(1+z).
\end{equation}

\noindent We cannot perform a true $K$-correction because most objects lack sufficient spectroscopic coverage. These equations will introduce systematic errors on the order of 0.1\,mag.

\begin{figure}[!htb]
    \centering
    \includegraphics[width=\columnwidth]{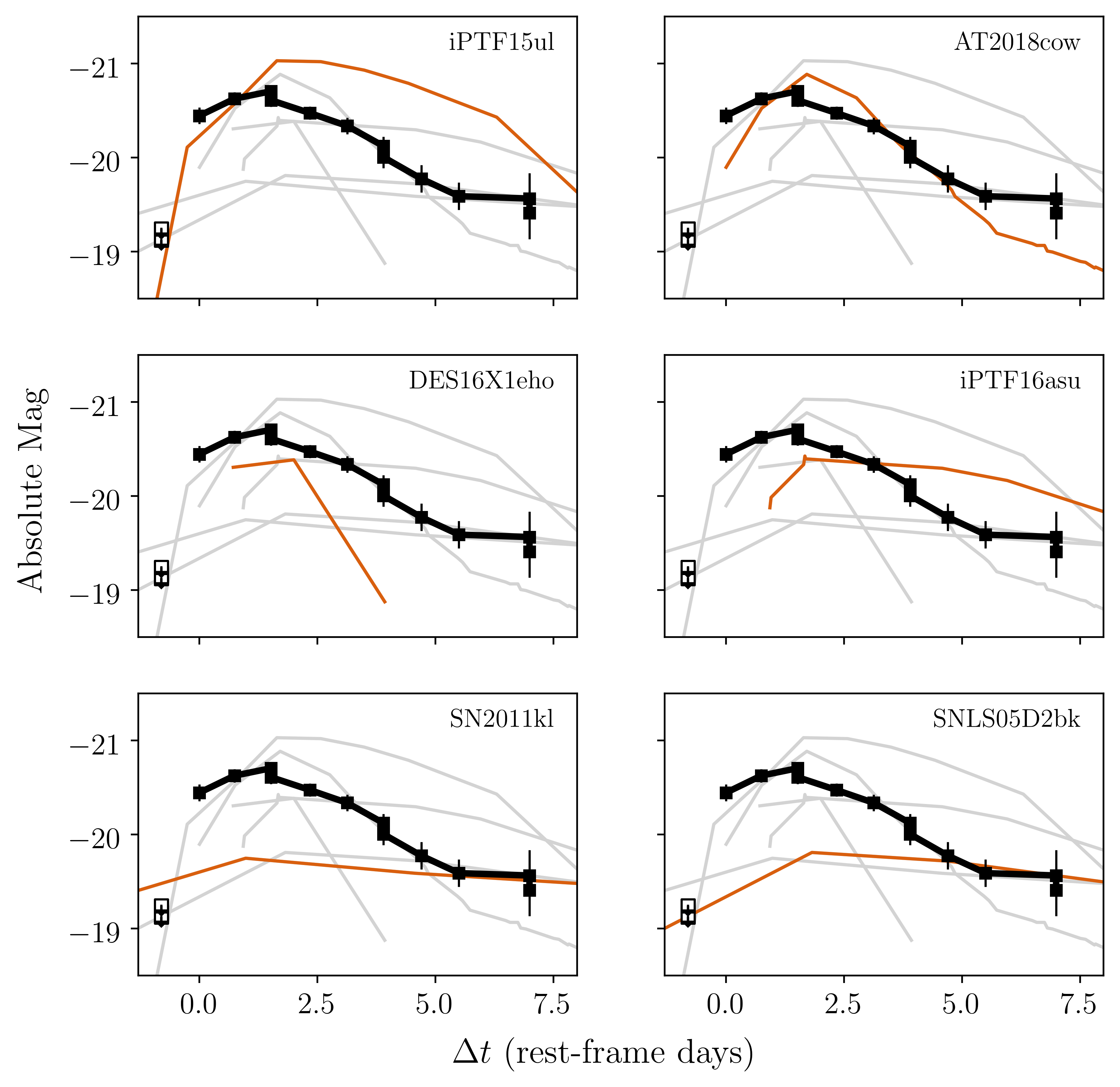}
    \caption{The rest-frame $g$-band (observer-frame $r$-band) light curve of \name\ (black line), compared to light curves of other transients in the literature
    in as close to the same rest-frame filter as possible.
    Each panel shows one transient highlighted in orange for comparison, with the rest shown in grey in the background.
    }
    \label{fig:optlc-comparison}
\end{figure}

In Figure~\ref{fig:optlc-comparison} we show the rest-frame $g$-band light curve of \name\ compared to the light curves of transients in Table~\ref{tab:literature}.
The fast rise time of \name\ is most similar to that of iPTF15ul, AT2018cow, and perhaps iPTF16asu: it is faster than SN\,2011kl and the SNLS transients.
\name\ fades much more quickly than iPTF16asu (which spectroscopically evolved into a Ic-BL SN) and in this sense is more similar to iPTF15ul and AT2018cow.
In terms of peak luminosity, \name\ is close to iPTF15ul, AT2018cow, DES16X1eho, and iPTF16asu,
and brighter than SN\,2011kl and the SNLS transients.
However, we caution that the high peak luminosity of iPTF15ul results from a large host-galaxy extinction inferred in
\citet{Hosseinzadeh2017},
without which the peak magnitude would be $-19.6$.

Next we consider color evolution.
\name\ showed tentative evidence for reddening over time, from $g-r=-0.47\pm0.09\,$mag at peak to $g-r=-0.03\pm0.21\,$mag in the final epoch a week later;
however, this is only a 2$\sigma$ change.
AT2018cow, iPTF15ul, and DES16X1eho remained very blue throughout the evolution of their optical light curves, whereas iPTF16asu reddened significantly as the SN became the dominant component.

Finally, we consider the spectral evolution of the transients in Table \ref{tab:literature}.
Peak-light spectra were not obtained for DES16X1eho \citep{Pursiainen2018} or the SNLS transients \citep{Arcavi2016}.
The peak-light spectra of iPTF16asu, AT2018cow, and SN\,2011kl were featureless \citep{Whitesides2017,Perley2019cow,Greiner2015},
and iPTF15ul\footnote{iPTF15ul was classified as Type Ibn in \citet{Hosseinzadeh2017}, but the lack of distinct \ion{He}{1} at peak make this classification uncertain.)} had a weak emission feature attributed to \ion{C}{3} \citep{Hosseinzadeh2017}.
After peak, iPTF16asu developed features of a Ic-BL SN \citep{Whitesides2017},
and AT2018cow had a complex spectral evolution, with a broad feature ($v>0.1c$) that appeared and disappeared over several days following peak light and a variety of emission lines that appeared one week later \citep{Perley2019cow}.
Unfortunately we do not have any spectra of \name\ after peak.

\subsection{Radio Light Curve}
\label{sec:comparison-radio}

In the previous section
(\S \ref{sec:comparison-optical})
we compared the optical properties of \name\ to the transients in Table~\ref{tab:literature}:
the light curve shape, the color evolution, and the spectrum.
In this section we compare the radio properties of \name\ to the same set of transients.

\begin{figure}[htb!]
    \centering
    \includegraphics[width=\columnwidth]{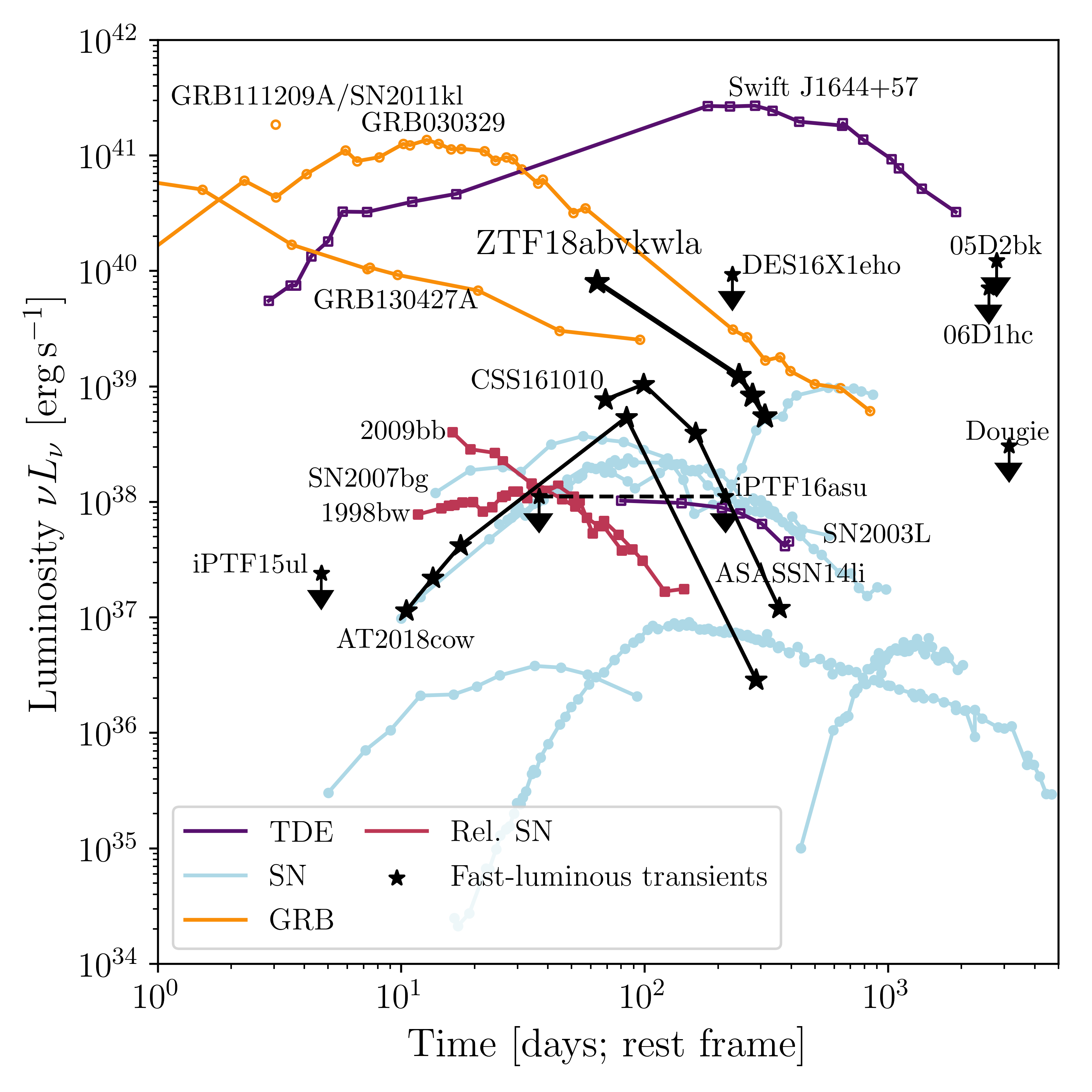}
    \caption{The 10\,\ghz\ radio light curve of \name\ compared to low-frequency (1--10\,\ghz) light curves of different classes of energetic explosions: tidal disruption events (purple; \citealt{Zauderer2011,Berger2012,Zauderer2013,Alexander2016,Eftekhari2018}), supernovae exploding in dense CSM (blue lines, $\gtrsim10^{37}\,\erg\,\psec$; \citealt{Soderberg2005,Soderberg2006,Salas2013}), relativistic Ic-BL supernovae (red lines; \citealt{Kulkarni1998,Soderberg2010}), AT2018cow (black line, small stars), long-duration gamma-ray bursts (orange lines; \citealt{Berger2003,Hancock2012,Perley2014,vanderHorst2014}), and ``ordinary'' supernovae ($\lesssim10^{37}\,\erg\,\psec$; \citealt{Weiler1986,Weiler2007,Horesh2013,Krauss2012}).
    The CSS161010 light curve was taken from \citet{Coppejans2020}.
    The AT2018cow light curve is at 9\,\ghz\ with data
    taken from \citet{Ho2019a}, \citet{Margutti2019},
    and \citet{Bietenholz2020}.
    \label{fig:radiolc-comparison}}
\end{figure}

Of the transients in Table~\ref{tab:literature},
only AT2018cow and GRB\,111209A/SN\,2011kl had a detected radio counterpart.\footnote{In the case of GRB\,111209A/SN\,2011kl, the radio emission was likely from the GRB afterglow itself \citep{Kann2018}.}
Prompt radio follow-up observations were also obtained for iPTF15ul\footnote{Observations of iPTF15ul were obtained within five days of the optical discovery, two observer-frame days after peak optical light, at 6\,\ghz\ and 22\,\ghz\ with the VLA,
at 15\,\ghz\ with the Arcminute Microkelvin Imager \citep{Zwart2008},
and at 95\,\ghz\ with the Combined Array for Research in Millimeter-wave Astronomy \citep{Bock2006}.
There was no detection at any frequency, 
with an RMS of 0.235\,mJy with CARMA and an RMS of 0.03\,mJy with AMI.} and iPTF16asu,
but neither was detected.
To our knowledge, Dougie, the SNLS transients, and DES16X1eho did not have deliberate radio follow-up observations;
we searched the VLASS archive and found that all except SNLS04D4ec were observed but none were detected.
In Figure~\ref{fig:radiolc-comparison} we show the radio measurements of the Table~\ref{tab:literature} transients compared to stellar explosions and tidal disruption events.
For completeness, we also searched the positions of all of the transients in the two largest collections of unclassified fast-rising luminous optical transients reported to date, PS1 \citep{Drout2014} and the Dark Energy Survey \citep{Pursiainen2018}.
None were detected, and the limits are listed in Table~\ref{tab:vlass}.

\begin{deluxetable}{lrrrrr}
\tablecaption{Radio limits for rapidly evolving transients in \citet{Drout2014} and \citet{Pursiainen2018} The $\Delta t$ is the number of days between the discovery date as listed in \citet{Drout2014} or the time of peak as listed in \citet{Pursiainen2018} and the epoch of the VLASS observation of that field.
\label{tab:vlass}
}
\tablehead{
\colhead{ID} & \colhead{$z$} & \colhead{RA} & \colhead{Dec} & \colhead{$\Delta t$} & \colhead{Limit} \\ 
& & [hh:mm:ss] & [dd:mm:ss] & (days) & ($\mu$Jy) \\ 
 }
\startdata
PS1-10ah & 0.074 & 10:48:15.784 & +57:24:19.48 & 2836 & 102 \\ 
PS1-11qr & 0.324 & 09:56:41.767 & +01:53:38.25 & 2467 & 130 \\ 
PS1-12bb & 0.101 & 09:57:23.866 & +03:11:04.47 & 2174 & 149 \\ 
PS1-12bv & 0.405 & 12:25:34.602 & +46:41:26.97 & 2642 & 129 \\ 
PS1-12brf & 0.275 & 22:16:06.892 & -00:58:09.81 & 1892 & 124 \\ 
PS1-11bbq & 0.646 & 08:42:34.733 & +42:55:49.61 & 2731 & 159 \\ 
PS1-13duy & 0.27 & 22:21:47.929 & -00:14:34.94 & 1505 & 127 \\ 
PS1-13dwm & 0.245 & 22:20:12.081 & +00:56:22.35 & 1422 & 155 \\ 
PS1-10iu & -- & 16:11:34.886 & +55:08:47.91 & 2689 & 103 \\ 
PS1-13aea & -- & 12:18:14.320 & +47:20:12.60 & 2199 & 88 \\ 
PS1-13bit & -- & 16:12:00.765 & +54:16:08.16 & 1618 & 104 \\ 
PS1-13cgt & -- & 16:18:56.245 & +54:19:33.71 & 1552 & 123 \\ 
DES15S1fli & 0.45 & 02:52:45.15 & -00:53:10.21 & 826 & 150 \\ 
DES13X3gms & 0.65 & 02:23:12.27 & -04:29:38.35 & 1520 & 139 \\ 
DES15S1fll & 0.23 & 02:51:09.36 & -00:11:48.71 & 826 & 139 \\ 
DES14S2anq & 0.05 & 02:45:06.67 & -00:44:42.77 & 1199 & 118 \\ 
DES14X3pkl & 0.3 & 02:28:50.64 & -04:48:26.44 & 1100 & 105 \\ 
DES15C3lpq & 0.61 & 03:30:50.89 & -28:36:47.08 & 849 & 145 \\ 
DES16S1dxu & 0.14 & 02:50:43.53 & -00:42:33.29 & 385 & 154 \\ 
DES15C3mgq & 0.23 & 03:31:04.56 & -28:12:31.74 & 835 & 99 \\ 
DES16X1eho & 0.76 & 02:21:22.87 & -04:31:32.64 & 365 & 152 \\ 
DES16X3cxn & 0.58 & 02:27:19.32 & -04:57:04.27 & 393 & 128 \\ 
DES15C3lzm & 0.33 & 03:28:41.86 & -28:13:54.96 & 839 & 106 \\ 
DES13C3bcok & 0.35 & 03:32:06.47 & -28:37:29.70 & 1513 & 98 \\ 
DES15C3nat & 0.84 & 03:31:32.44 & -28:43:25.06 & 810 & 108 \\ 
DES15C3opk & 0.57 & 03:26:38.76 & -28:20:50.12 & 777 & 125 \\ 
DES15C3opp & 0.44 & 03:26:57.53 & -28:06:53.61 & 781 & 112 \\ 
DES13X3npb & 0.5 & 02:26:34.11 & -04:08:01.96 & 1411 & 122 \\ 
DES16C3axz & 0.23 & 03:31:14.15 & -28:40:00.25 & 523 & 100 \\ 
DES16C3gin & 0.35 & 03:31:03.06 & -28:17:30.98 & 391 & 107 \\ 
DES14X1bnh & 0.83 & 02:14:59.79 & -04:47:33.32 & 1172 & 145 \\ 
DES16X3ega & 0.26 & 02:28:23.71 & -04:46:36.18 & 357 & 111 \\ 
DES15C3mfu & -- & 03:28:36.08 & -28:44:20.00 & 835 & 187 \\ 
DES13C3abtt & -- & 03:30:28.91 & -28:09:42.12 & 1513 & 107 \\ 
DES15C3pbi & -- & 03:28:56.68 & -28:00:07.98 & 772 & 182 \\ 
DES15X3atd & -- & 02:23:21.64 & -04:17:28.95 & 830 & 146 \\ 
DES13C3nxi & -- & 03:27:51.22 & -28:21:26.21 & 1559 & 75 \\ 
DES13C3smn & -- & 03:27:53.08 & -28:05:00.93 & 1564 & 124 \\ 
DES13X3aakf & -- & 02:22:50.84 & -04:41:57.01 & 1441 & 108 \\ 
DES13X3afjd & -- & 02:28:00.31 & -04:34:59.39 & 1411 & 123 \\ 
DES13X3kgm & -- & 02:26:00.92 & -04:51:59.29 & 1508 & 103 \\ 
DES16S2fqu & -- & 02:47:05.94 & -00:20:50.40 & 356 & 139 \\ 
DES16X1ddm & -- & 02:15:18.88 & -04:21:52.07 & 386 & 111 \\ 
DES16X3ddi & -- & 02:21:45.39 & -04:41:08.95 & 393 & 127 \\ 
DES16X3erw & -- & 02:24:49.31 & -04:30:51.45 & 357 & 117 \\
\enddata
\end{deluxetable}

As shown in Figure \ref{fig:radiolc-comparison},
ZTF18abvkwla is most similar in luminosity
to long-duration GRB afterglows \citep{Berger2003,Perley2014}.
The SED is also similar:
in \S\ref{sec:radio-obs} we found that the SED of ZTF18abvkwla peaked near 10\,\ghz\ at $\Delta t=81\,\days$,
while the SED of GRB\,030329 ($z=0.1685$) peaked at 5\,\ghz\ \citep{Berger2003} at 67 days post-explosion,
and the SED of GRB\,130427A ($z=0.340$) peaked at 10\,\ghz\ \citep{Perley2014} at a similar epoch post-explosion.

\subsection{A Starburst Host Galaxy}
\label{sec:comparison-host}

In Sections \ref{sec:comparison-optical} and \ref{sec:comparison-radio} we compared the optical and radio properties of \name, respectively, to other transients in the literature.
Here we put its host galaxy properties into context.

Galaxies with very high specific star-formation rates (e.g., sSFR $\gtrsim 10^{-8}$ yr$^{-1}$, our operational definition of a ``starburst'') contribute a small fraction of star-formation in the low-redshift Universe \citep{Lee2009}, so the appearance of \name\ in such a galaxy (sSFR $\sim$ $1.4\times10^{-8}$ yr$^{-1}$) is notable.
However, their contribution to low-metallicity star-formation is more significant, as they are typically low-mass
and therefore low-metallicity \citep{Tremonti2004}.
They are also promising candidates to experience a top-heavy IMF \citep{Dabringhausen2009} and potential sites of enhanced binary or dynamical stellar interactions \citep{vandenHeuvel2013}. Each of these mechanisms have been appealed to in attempts to interpret the relatively high abundance of exotic transients of other types found in these systems, including superluminous SNe (SLSNe; \citealt{Lunnan2014,Leloudas2015,Perley2016,Schulze2018}), broad-lined Ic SNe \citep{Modjaz2019}, GRBs \citep{Fruchter2006,Kruhler2015,Schulze2015,Vergani2015}, and at least some fast radio bursts \citep{Katz2016,Tendulkar2017}.

\begin{figure*}[htb!]
         \begin{minipage}[l]{1.0\columnwidth}
             \centering
             \includegraphics[width=0.9\columnwidth]{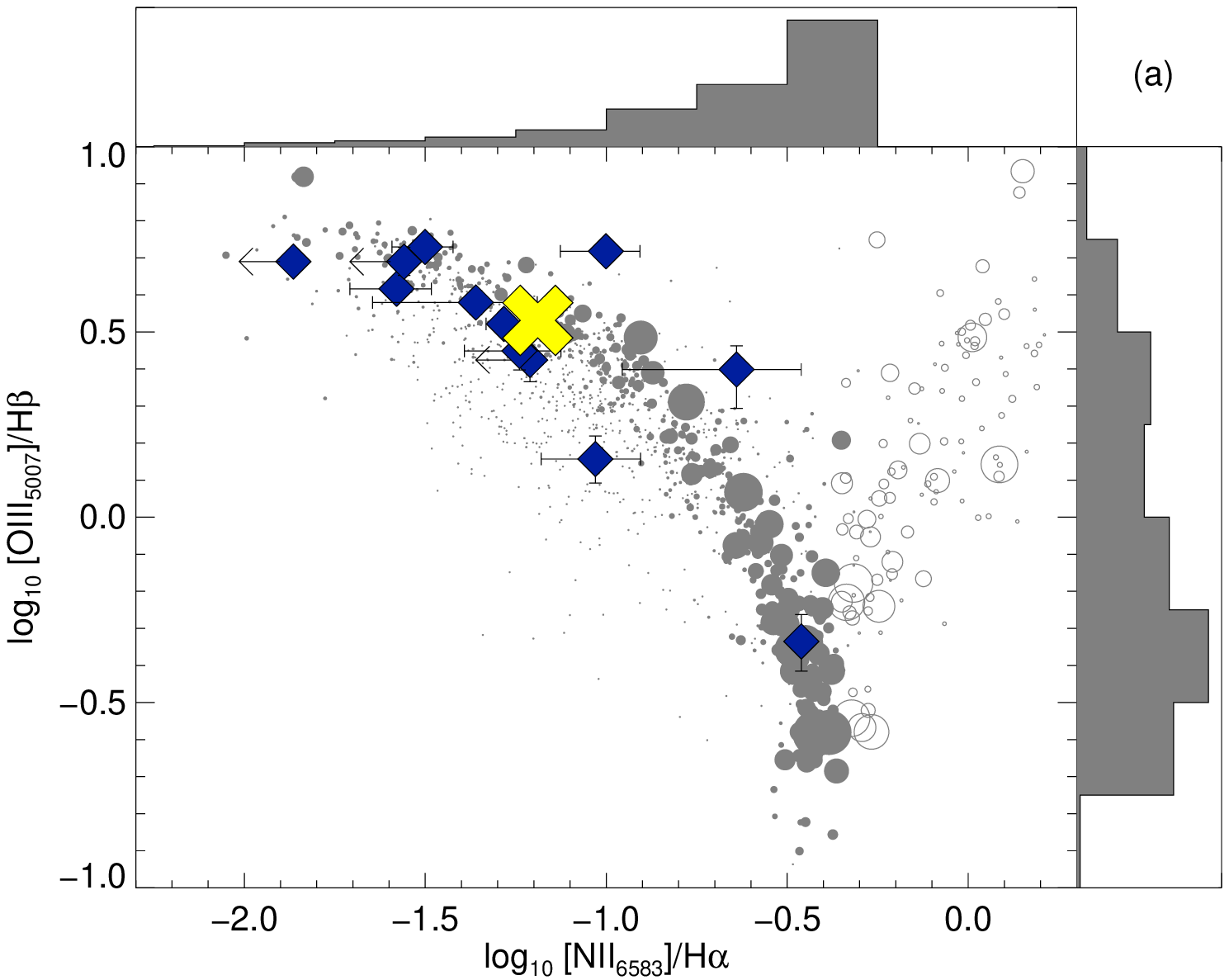}
         \end{minipage}
         \hfill{}
         \begin{minipage}[r]{1.0\columnwidth}
             \centering
             \includegraphics[width=0.9\columnwidth]{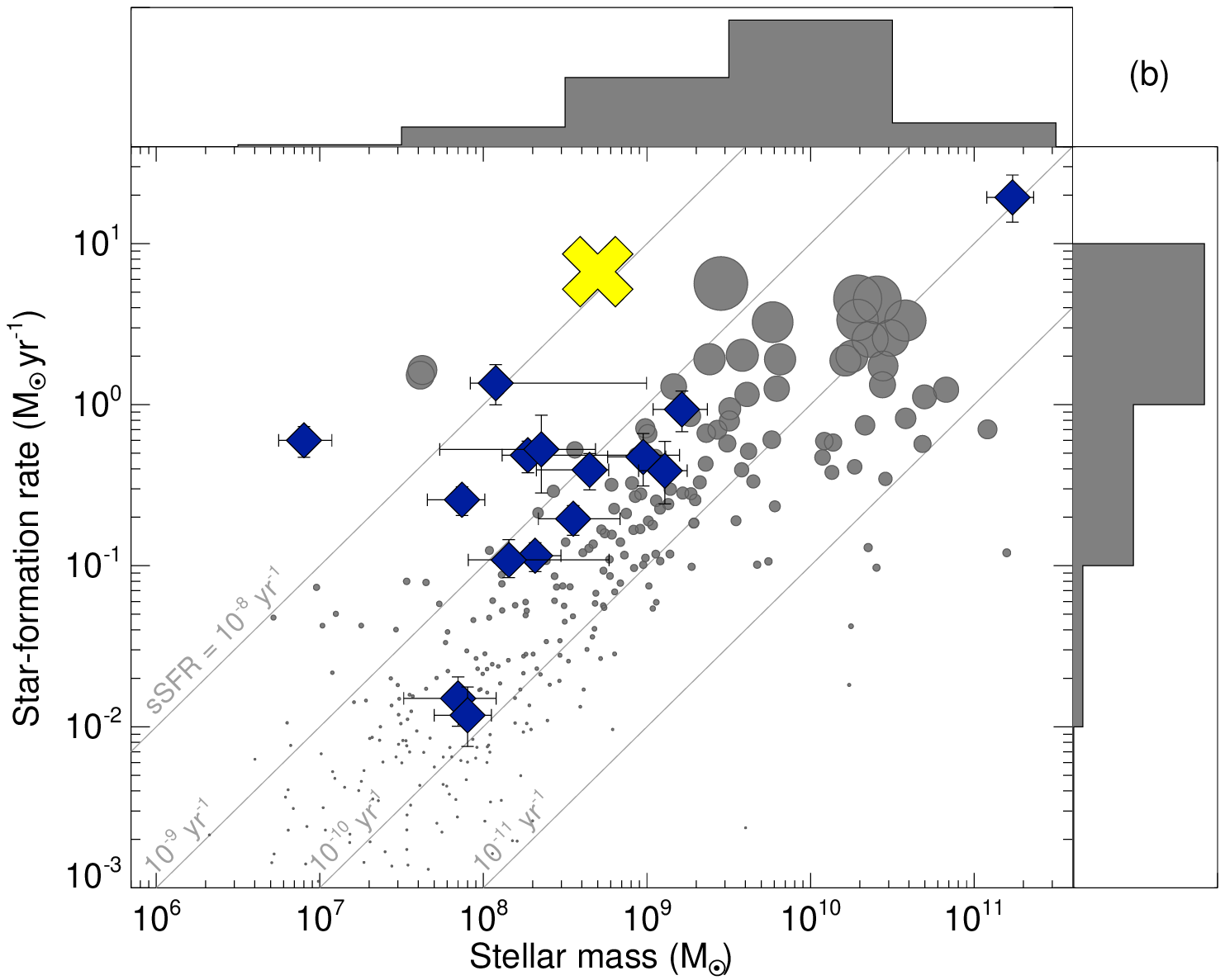}
         \end{minipage}
         \begin{minipage}[l]{1.0\columnwidth}
             \centering
             \includegraphics[width=0.9\columnwidth]{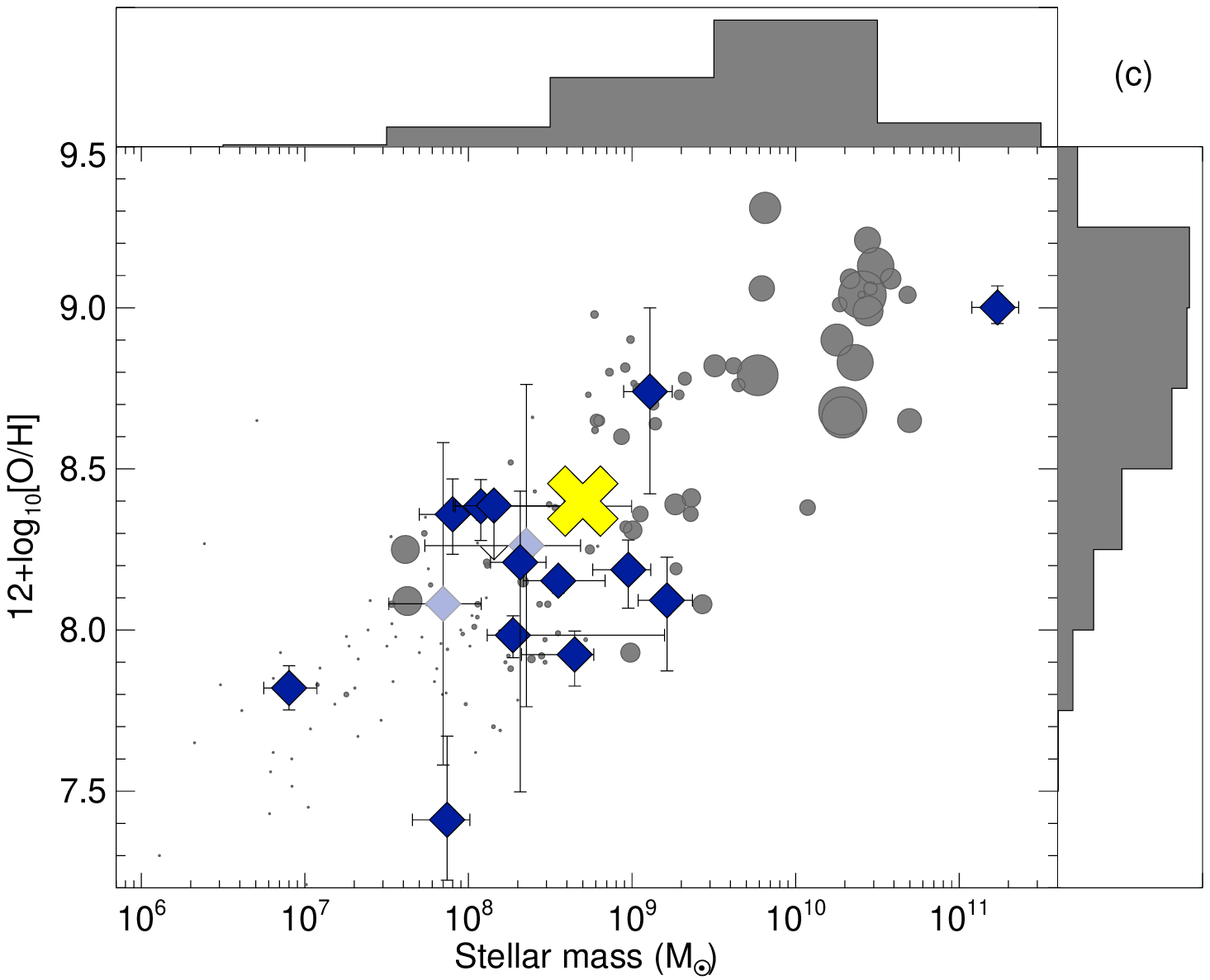}
         \end{minipage}
         \hfill{}
         \begin{minipage}[r]{1.0\columnwidth}
             \centering
             \includegraphics[width=0.9\columnwidth]{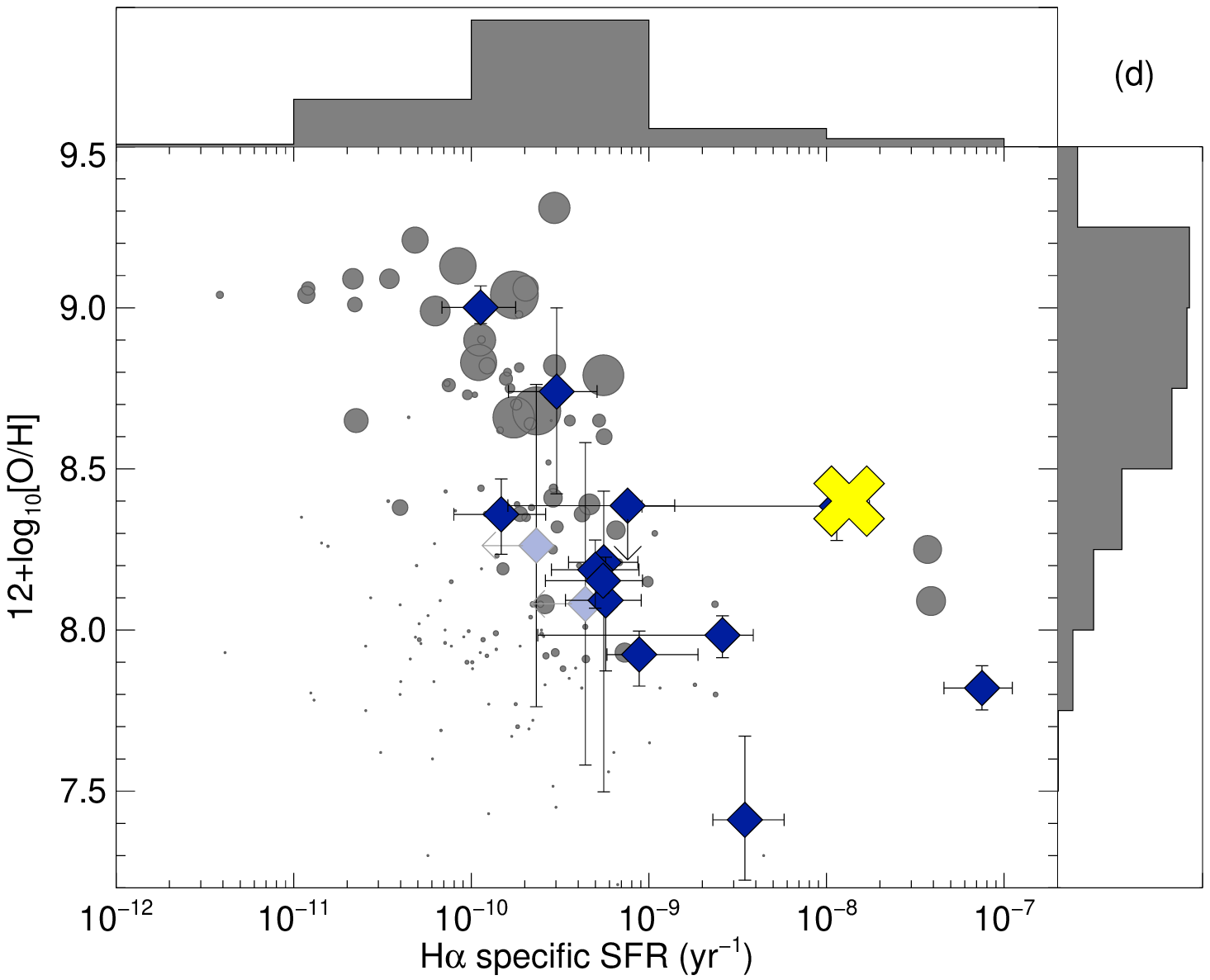}
         \end{minipage}
    \caption{Comparison of the host galaxy of \name\ to $<$11\,Mpc comparison galaxies (grey) and to the host galaxies of nearby hydrogen-poor SLSNe (diamonds), as in \citealt{Perley2016}.
    Light diamonds indicate mass-metallicity estimated metallicities. Comparison galaxies are weighted by their SFR; histograms show the SFR-weighted binned totals on each axis. \name\ is indicated by a yellow cross. From top left:  (a) BPT diagram.  (b) Mass--star-formation rate relation.  (c) Mass--metallicity relation.  (d) Specific star-formation-rate--metallicity relation.  The host is a starbursting galaxy with no evidence of AGN activity, and while it is metal-poor it is not particularly so given its mass.}
    \label{fig:hostplots}
    \end{figure*}

Based on our measurements in \S \ref{sec:obs-opt-spec} we conclude the following about the host of \name:

\emph{The host is not an AGN} --- We confirm the lack of any evidence for an optical AGN based on the very weak [NII] emission.  The host falls squarely in the star-forming locus of the BPT diagram (Figure~\ref{fig:hostplots}a).

\emph{The host metallicity is typical for its mass} --- The host is relatively metal-poor: the precise number is of course scale-dependent, but using the Z94 scale we calculate [O/H] of 8.45, or about 0.6$\times$Solar.  This is a lower metallicity than the majority of star-formation in the local Universe, but not an outlier and unexceptional for low-mass galaxies in particular (Figure~\ref{fig:hostplots}b).

\emph{The star-formation intensity is similar to extreme SLSN and GRB hosts} --- The most striking nature of the host galaxy is its very high specific star-formation rate, which is evident in Figure~\ref{fig:hostplots}c and \ref{fig:hostplots}d.
 
The host of AT\,2018cow was also a dwarf galaxy, although it was more massive than that of \name\ and not starbursting,
with a mass and star-formation rate of $1.4 \times 10^{9}\,\msol$ and $0.22\,\msol\,\pyr$ respectively \citep{Perley2019cow}.
The host galaxy of DES16X1eho had a stellar mass $\log(M/\msol)=9.96^{+0.14}_{-0.51}$ and a specific SFR of $\log(\mathrm{sSFR/\msol\,\pyr})=-9.25$ \citep{Pursiainen2018}.
The host galaxy of iPTF16asu had a stellar mass $M=4.6^{+2.0}_{-2.3}\times10^{8}\,\msol$
and an H$\alpha$ SFR of 0.7\,\msol\,\pyr,
corresponding to a sSFR of 1.4\,\pgyr \citep{Whitesides2017}.
Finally, the host galaxies of the SNLS transients harbored relatively evolved stellar populations,
and were noted to be markedly different from starburst galaxies \citep{Arcavi2016}.

\section{Interpretation}
\label{sec:interpretation}

Even with the small number of events in the Table~\ref{tab:literature} menagerie,
the diversity of optical and radio properties (\S \ref{sec:comparison-optical}, \S \ref{sec:comparison-radio}) suggests that there are several progenitor systems involved.
In this section we model the optical and radio light curves of ZTF18abvkwla and discuss the implications for the progenitor.

\subsection{Modeling the Optical Light Curve}
\label{sec:lc-modeling}

Shock-interaction with extended low-mass material is an efficient mechanism for producing a fast-peaking luminous optical light curve.
Shock breakout occurs when the photon diffusion time drops below the shock crossing time ($\tau < c/v_s$, where $\tau$ is the optical depth and $v_s$ is the shock velocity).
For normal stellar progenitors, this emission is primarily at X-ray and UV wavelengths and lasts for seconds to a fraction of an hour.
In the wake of this shockwave, the outer stellar material is heated to high temperatures, and as it cools it radiates on the timescale of a day (``cooling envelope'' emission).
See \citet{Waxman2017} for a review.

Prior to core-collapse, massive stars can undergo mass-loss via steady winds or eruptive episodes \citep{Smith2014}.
As a result, a star can be surrounded by dense, recently-expelled material at the time of explosion.
If this material is optically thick,
it increases the effective radius of the star
and prolongs the light curve from shock breakout.
If the light curve of \name\ arises from shock breakout in a shell, we can estimate the radius of this extended material (CSM) assuming a rise to peak bolometric luminosity $\trise<2\,\days$, a peak luminosity $L_\mathrm{bol}>10^{44}\,\erg\,\psec$ and a typical SN shock velocity of $10^4\,\km\,\psec$.
The rise timescale is

\begin{equation}
\begin{split}
    t_{\mathrm{BO}} & \sim \frac{R_\mathrm{CSM}}{v_s} \\
    & = (1.3\,\days)
    \left( \frac{R_\mathrm{CSM}}{10^{15}\,\cm} \right) 
    \left( \frac{v_s}{10^4\,\km\,\psec} \right)^{-1}.
\end{split}
\end{equation}

\noindent For \name, we find $R_\mathrm{CSM}<1.5 \times 10^{15}\,\cm$.

We can also estimate the mass in the shell, assuming that the shock deposits half its kinetic energy $(1/2)\rho v_s^2$ and that this deposited energy is $E_\mathrm{BO}\sim4\pi R^2dR e_s$ where
the energy density reflects the amount of thermal energy in the layer. The luminosity scales as

\begin{equation}
\begin{split}
    L_\mathrm{BO} & \sim \frac{E_\mathrm{BO}}{t_\mathrm{cross}} \sim \frac{v_s^3}{4} \frac{dM}{dR}
    = (2.2 \times 10^{45}\,\erg\,\psec) \\
    & \times \left( \frac{v_s}{10^4\,\km\,\psec} \right)^3
    \left( \frac{dM}{\msol} \right)
    \left( \frac{dR}{10^{15}\,\cm} \right)^{-1}.
\end{split}
\end{equation}

\noindent Assuming $dR\sim R$, we find $M_\mathrm{CSM}<0.07\,\msol$.
In this framework, the differences in the light curves of different objects corresponds to differences in the shell mass, shell radius, and shock velocity.
The luminosity is most sensitive to the velocity, so it is possible that the transients in Table~\ref{tab:literature} are distinguished by fast velocities, which would naturally explain the inclusion of a Ic-BL SN.
For a fixed shock velocity, a fast rise time corresponds to a small shell radius, which in turn requires a large shell mass to produce a high luminosity.

Another possibility is that the light curve is powered not by shock breakout in a shell,
but by post-shock envelope-cooling emission.
For example, this was the model invoked for iPTF16asu \citep{Whitesides2017},
which led to an inferred shell mass of 0.45\,\msol\ and a shell radius of $1.7 \times 10^{12}\,\cm$.
The light curve of \name\ has a similar rise time but a higher peak luminosity than that of iPTF16asu,
and the effective temperature at peak is significantly higher.
According to the one-zone analytic formalism in \citet{Nakar2014} and \citet{Piro2015},
a higher peak temperature for a fixed rise time and a fixed opacity arises from a larger shell radius.
A larger shell radius can also explain the higher bolometric luminosity, although that could also arise from a larger explosion energy or faster ejecta velocity.

Another mechanism suggested to explain the optical light curve of AT2018cow was reprocessing by dense outer ejecta \citep{Margutti2019}.
In this picture,
a central source (such as an accretion disk or magnetar) emits high-energy (i.e. X-ray) emission, which is reprocessed by
surrounding material to produce lower-energy (i.e. optical) radiation. 
This is one setup for tidal disruption events, in which case the surrounding material is unbound stellar debris \citep{Strubbe2009}.
Indeed, several properties of \name\ and AT2018cow are similar to TDEs in the literature, such as
the photospheric radius of $10^{14}$--$10^{15}\,\cm$,
the effective temperature of $10^{4}$\,K, and high radio luminosities attributed to jets (for reviews of TDE observations, see \citet{Gezari2012} and \citet{Komossa2015}).

Regardless of the power source at peak,
we also use the optical light curve to put an upper limit on the mass of \nickel\ that could have been synthesized in the explosion. Using Equation (16) in \citet{Kasen2017}, the luminosity from the radioactive decay of \nickel\ is

\begin{equation}
\begin{split}
    L (t) & =  2 \times 10^{43} \left( \frac{\mni}{\msol} \right) \\ 
    & \times \left[ 3.9 e^{-t/\tau_{\mathrm{Ni}}} + 0.678 \left( e^{-t/\tau_{\mathrm{Co}}} - e^{-t/\tau_{\mathrm{Ni}}} \right) \right]
    \erg\,\psec
\end{split}
\end{equation}

\noindent where $\tau_\mathrm{Ni}=8.8\,$d and $\tau_\mathrm{Co}=113.6\,$d.
Using the final $g$-band measurement
($g=21.51\pm0.21$) at $\Delta t=10\,\days$
($\Delta t=8\,\days$ rest-frame)
$L \approx \lambda F_\lambda \approx 1.4 \times 10^{43}\,\erg\,\psec$, so the amount
of \nickel\ that could power the light curve at this epoch is $\mni \lesssim 0.36\,\msol$ (Figure~\ref{fig:lightcurve}).
From a compilation of CC SNe, \citet{Lyman2016}
found nickel masses of $0.11\pm0.04\,\msol$ for Type IIb SNe,
$0.17\pm0.16\,\msol$ for Type Ib SNe,
$0.22\pm0.16\,\msol$ for Type Ic SNe,
and $0.32\pm0.15\,\msol$ for Type Ic-BL SNe.
So, we cannot rule out an underlying nickel-powered light curve for ZTF18abvkwla.

\subsection{Modeling the Radio Light Curve}
\label{sec:modeling-radio}

The high luminosity and fast variability timescale of the 10\,\ghz\ light curve implies a high brightness temperature $T_B \approx 10^{11}\,$K,
so we conclude that the emission is synchrotron radiation.
In the first epoch, the 10\,\ghz\ observation is declining and has an in-band (8--12\,\ghz) spectral index of
$\alpha=-0.16 \pm 0.05$ where $F_\nu \propto \nu^{-\alpha}$.
This is much shallower than the optically thick ($\alpha=-2.5$) or the optically thin ($\alpha=+0.7$) regimes of a synchrotron self-absorption (SSA) spectrum,
which suggests that the peak of the SED is near 10\,\ghz\ (observer-frame) at this epoch.
In what follows,
we assume that the SSA spectrum has a rest-frame peak frequency $\nu_p \lesssim 8\,\ghz$ (the bottom of the band) and a rest-frame peak flux density $F_p \gtrsim 0.364\,$mJy.

\begin{figure}[htb!]
\includegraphics[width=\linewidth]{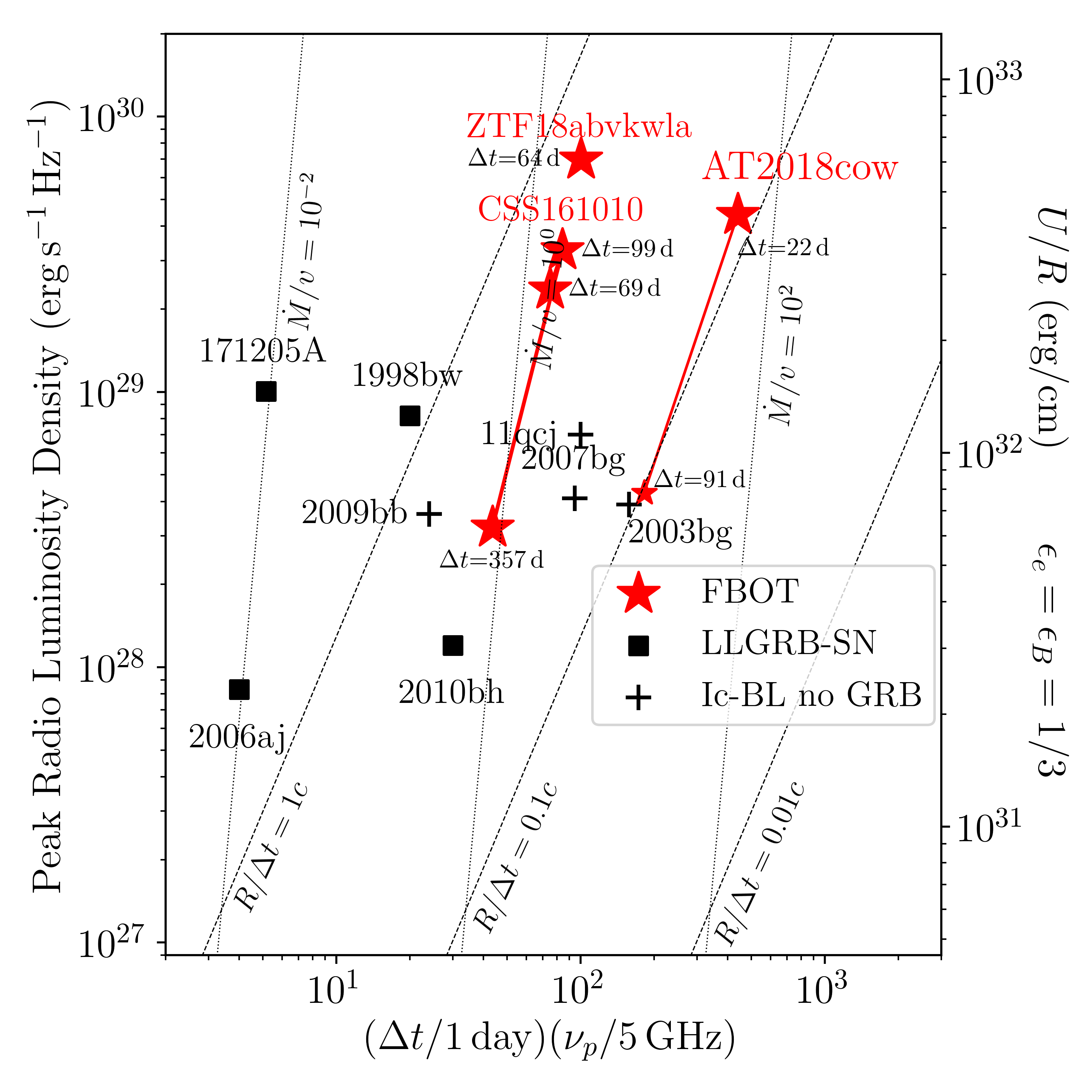}
\caption{Approximate luminosity and frequency of the SSA peak of ZTF18abvkwla at $\Delta t=81\,\days$ (observer-frame), compared to other energetic explosions in the literature, including AT2018cow \citep{Ho2019a,Margutti2019} and CSS161010 \citep{Coppejans2020}. Lines of constant mass-loss rate are shown in units of $10^{-4}\,\msol\,\pyr$ scaled to a wind velocity of 1000\,\km\,\psec.
The corresponding energy of the explosion (assuming equipartition) is shown on the right-hand side.
\label{fig:lum-tnu}}
\end{figure}

When the SSA peak is known,
the outer shock radius $R_p$ and magnetic field strength $B_p$ can be derived assuming that energy is equally partitioned into magnetic fields and relativistic electrons \citep{Scott1977,Readhead1994}.
We use the equations for $R_p$ and $B_p$ for radio SNe in \citet{Chevalier1998} (Equations 11 and 12).
Assuming an optically thin spectral index of $\nu^{-1}$ and a filling factor $f=0.5$, we find $R_p \gtrsim 8.0 \times 10^{16}\,\cm$ and $B_p \lesssim 0.51\,$G.
So, the mean velocity until $t_\mathrm{obs}=81\,\days$ is
$\Gamma \beta c= R_p (1+z)/t_\mathrm{obs} = 0.38c$.
Using Equations 12, 16, and 23 in \citet{Ho2019a},
and assuming $\epsilon_e = \epsilon_B = 1/3$,
we find that the shock has swept up energy $U = 3.4 \times 10^{49}\,\erg$ into an ambient medium of density $n_e=190\,\pcmcub$, corresponding to a mass-loss rate of $\dot{M} = 5.8 \times 10^{-4}\,\msol\,\pyr$ assuming a wind velocity $v_w = 1000\,\km\,\psec$.
In Figure \ref{fig:lum-tnu} we show these quantities compared to those of other energetic explosions.
The peak radio luminosity density is directly proportional to $U/R$, the energy swept up by the shock divided by the shock radius (right-hand side of Figure \ref{fig:lum-tnu}).
So, the fact that ZTF18abvkwla, AT2018cow, and CSS161010 are distinguished by high radio luminosities is primarily a consequence of a large explosion energy.

\subsection{Progenitor Systems and a Search for an Associated Gamma-ray Burst}
\label{sec:progenitors}

The physical setups outlined in \S \ref{sec:lc-modeling} --- a shock driven through a shell, reprocessing of a high-energy compact source by optically thick material --- could arise in a variety of different progenitor systems.
An additional clue for \name\ is the host galaxy, which experienced a very recent burst of star-formation activity.
In that sense, a massive-star origin seems most natural.

AT2018cow was suggested to have two distinct components:
a shock driven through dense equatorial material (producing the optical emission),
and a faster polar outflow (producing the radio emission; \citealt{Margutti2019}).
As shown by early millimeter observations \citep{Ho2019a}, later radio observations \citep{Margutti2019},
and Very Long Baseline Interferometry \citep{Bietenholz2020,Mohan2020},
the fast outflow was subrelativistic with a near-constant velocity of $v=0.1c$.
In ZTF18abvkwla,
the radio-emitting ejecta is faster:
$>0.38c$ at the same epochs when the outflow from AT2018cow was $0.1c$.
As shown in Figure \ref{fig:lum-tnu},
the higher luminosity at late times arises from this faster velocity; the explosion energy of the two events appears to have been similar.

Because the late-time radio light curve is similar of that of GRBs,
we searched for potential GRB counterparts to \name\ in the period between the last non-detection (MJD 58372.4206; 2018-09-11 10:05:39.84) and the first detection (MJD 58373.4075; 2018-09-12 09:46:48.00).
There were two bursts detected by the interplanetary network (IPN; \citealt{Hurley2010,Hurley2016}),
one by the Gamma-ray Burst Monitor (GBM) aboard the \emph{Fermi} spacecraft \citep{Gruber2014,Kienlin2014,Bhat2016} and one detected by the  Konus-\emph{Wind} experiment aboard the \emph{Wind} spacecraft \citep{Aptekar1995}.
The positions of both bursts are inconsistent with that 
of \name.

Due to the lack of detected GRB,
we can set a limit on the fluence and corresponding isotropic  equivalent energy of a prompt burst associated with \name. 
The IPN has essentially a 100\% duty cycle across the sky,
and detects GRBs with $E_\mathrm{p}>20\,\kev$ down to $6 \times 10^{-7}\,\erg\,\pcmsq$ at 50\% efficiency \citep{Hurley2010,Hurley2016}.
At $t_0$, the estimated 20--1500\,\kev\ limiting peak flux at the position of \name\ was $2 \times 10^{-7}\,\erg\,\pcmsq\,\psec$ for a Band model that has $E_\mathrm{pk}$ in the 50--500\,\kev\ range.
At the distance of \name,
this corresponds to a limit on the isotropic peak luminosity
of $L_\mathrm{iso}<5 \times 10^{49}\,\erg\,\psec$.
Therefore we strongly disfavor an on-axis classical GRB (which is also consistent with the lack of observed optical afterglow emission).

Among GRBs, two events have shown evidence for a luminous optical blackbody component at early times:
GRB\,060218 ($z=0.033$; \citealt{Soderberg2006,Mirabal2006,Pian2006,Sollerman2006,Ferrero2006}) and GRB\,101225A ($z\approx0.3$; \citealt{Thone2011}).
GRB\,060218 was a very long-duration ($T_{90}\approx2100\,$s) low-luminosity ($L_\mathrm{iso}=2.6 \times 10^{46}\,\erg\,\psec$) GRB associated with the Ic-BL SN\,2006aj \citep{Cano2017}.
A GRB with these properties cannot be ruled out by our limits.
GRB\,101225A also had a very long duration $T_{90}>2000\,\sec$,
and a candidate faint ($M\approx -16.7$) SN counterpart.

As in the case of AT2018cow,
we cannot rule out a TDE origin.
In that case, the similarity to the light curve of AT2018cow would suggest a similar kind of system, i.e. an intermediate-mass black hole ($M\sim10^4\,\msol$; \citealt{Perley2019cow}) with a white dwarf \citep{Kuin2019} or a Solar-type \citep{Perley2019cow} stellar companion.
In the case of AT2018cow, the main argument against a TDE hypothesis was the large ambient density ($10^{5}\,\pcmcub$) from millimeter \citep{Ho2019a} and radio \citep{Margutti2019} observations.
For \name, assuming that the flat spectral index indicates a 10\,\ghz\ peak at $81\,\days$, we find a much lower density ($10^{2}\,\pcmcub$).
Among TDEs,
the radio light curve of \name\ is most similar to that of the TDE candidate IGR J12580+0134 \citep{Irwin2015}, which had a nearly identical $\nu L_\nu$ (and fade rate) one year post-discovery.
The radio emission from IGR J12580+0134 has been attributed to an off-axis relativistic jet \citep{Irwin2015,Lei2016}
but interpretation is complicated by the coincidence of the source with a known AGN.

\section{Rate Estimate}
\label{sec:rates}

An important clue to the progenitor of sources like \name\ is the cosmological rate.
Furthermore, three fast-luminous transients -- SN\,2011kl (associated with GRB\,111209A), AT2018cow, and \name\ -- have detected luminous radio emission,
although the radio emission from SN\,2011kl likely arose from the GRB afterglow.
Clearly, being able to recognize additional members of this phase-space in optical surveys would be valuable for radio follow-up observations.
In this section, we conduct an archival search of 18 months of the 1DC survey (2018 Apr 3 -- 2019 Oct 18 UT) to estimate the rate of transients in the phase-space of Figure~\ref{fig:lum-rise}
and delineate false positives.

First we selected field-nights in the survey for which the 1-night coverage was approximately maintained.
Specifically, we require

\begin{itemize}
    \item at least one observation the night before ($0.5 < dt < 1.5\,$days)
    \item at least one observation two nights before ($2.5 < dt < 1.5\,$days)
    \item at least three observations in the next five nights ($dt<5.5\,$days)
\end{itemize}

We find 8064 fields satisfying these criteria.
Of these, 6854 fields (85\%) have limiting magnitude $>19.75\,$mag
and 4596 fields (57\%) have limiting magnitude $>20.5\,$mag.
The dominant effect is lunation, with some night-to-night variations due to weather.

For each of the 8064 field-nights,
we searched for fast transients.
To detect a fast transient, we require that the peak of the light curve be ``resolved:'' that is, that there are measurements both before and after peak light that are $>0.75$\,mag fainter than the peak magnitude.
We then measure the time from 0.75\,mag below peak to peak by linearly interpolating the light curve.
If this rise time is $<5$\,d, we include the transient in our sample.
More specifically, we filtered sources as summarized in Table~\ref{tab:filtering-steps}.
We scanned the remaining 659 sources by eye and
removed sources with very noisy light curves or flaring behavior.

\begin{deluxetable}{lr}[htb!]
\tablecaption{Filtering criteria for sources similar to \name\ in the ZTF 1DC survey\label{tab:filtering-steps}} 
\tablewidth{0pt} 
\tablehead{\colhead{Criteria} & \colhead{\# sources remaining}
}
\tabletypesize{\scriptsize} 
\startdata 
Real$^{a}$, bright$^{b}$, pos. sub.$^{c}$, not star$^{d}$ & 758,528 \\
Short duration$^{e}$ and peak resolved$^{f}$ & 659 \\
\enddata
\tablecomments{
$^{a}$ $\texttt{drb}>0.99$
$^{b}$ $\texttt{magpsf}<20$
$^{c}$ \texttt{isdiffpos}=`t' or '1'
$^{d}$ not($\texttt{sgscore1}>0.76$ and $\texttt{distpsnr1}<1$)
$^{e}$ Duration between 1 and 100 days
$^{f}$ Peak has preceding or subsequent detection/non-detection in a $\pm5\,\days$ window that is at least 0.75\,mag fainter
}
\end{deluxetable}

In Table~\ref{tab:ztf-fbots}
we list all 27 sources with rise times faster than 5\,\days, including \name\ itself.
Five sources are spectroscopically classified SNe: two Type II, two Type Ibn, and one Type IIb. Three sources are classified as CVs, two spectroscopically and one by cross-matching with the AAVSO International Variable Star Index VSX \citep{Watson2017}.
Two are very likely flare stars based on previous detections in Pan-STARRS individual-epoch images,
and a third is a likely flare star based on a GALEX counterpart.
Nine sources are likely extragalactic (based on proximity to a host galaxy).
When redshift estimates for these galaxies were not available, we attempted to obtain them using LRIS on 17 Feb 2020.
Two sources remain without definitive redshift estimates,
so we provide a photometric redshift from LegacySurvey DR8.
One source (ZTF18abxxeai) has a very faint host classified as a PSF in LegacySurvey DR8,
and the remaining five sources have no clear host counterpart.

Of the sources with a definitive host redshift measurement,
\name\ is the only one that is more luminous than $M=-20$\,mag.
Clearly, the primary interlopers in searches for transients like \name\ are CVs and less luminous SNe.
CVs can be ruled out on the basis of repeated flaring,
whereas less luminous SNe can only be ruled out if the redshift of the host galaxy is known \emph{a priori}.
Aside from \name, eight transients in our sample remain as possibly having $M_\mathrm{g,peak}<-20$,
although the lack of an obvious host for six of them suggest that these may be CVs.

\begin{deluxetable*}{lrrrrrr}[!ht]
\tablecaption{Fast-rising transients in ZTF resulting from our archival search of the one-day cadence survey. In the redshift column, a range refers to the 68 percentile range on the photometric redshift from LegacySurvey DR8 (we provide a corresponding range of absolute magnitude) and a single value corresponds to a spectroscopic redshift.
When the distance is known,
the peak mag is an absolute magnitude,
and when the distance is not known the peak mag is an apparent magnitude.
These values correspond to the filter as close to rest-frame $g$-band as possible,
and when the distance is not known they correspond to the observed
$g$-band filter.
Magnitudes are corrected for Galactic extinction
and timescales are in rest-frame when the redshift is known, and in observer-frame when the redshift is not known.
\label{tab:ztf-fbots}} 
\tablewidth{0pt} 
\tablehead{\colhead{ZTF Name (IAU Name)} & \colhead{Redshift} & \colhead{Peak Mag} & \colhead{\trise} & \colhead{\tfade} & \colhead{Type} & \colhead{Notes} \\
\colhead{} & \colhead{} & \colhead{} 
}
\tabletypesize{\scriptsize} 
\startdata 
18abvkwla & 0.2714 & $-20.59\pm0.07$ & $1.83\pm0.05$ & $3.12\pm0.22$ & FBOT & This paper \\
19aavbjfp (SN2019fkl) & 0.028 & $-17.4\pm0.4$ & $3.2\pm0.9$ & $21.8\pm6.1$ & SN\,II \\
19abgbdcp (AT2019lbv) & 0.0318 & $-18.36\pm0.03$ & $2.32\pm0.03$ & $14.4\pm0.9$ & SN\,II \\
18aalrxas & 0.0588 & $-18.43\pm0.03$ & $1.86\pm0.02$ & $2.4\pm0.3$ & SN\,IIb & \citet{Fremling2019} \\
19abuvqgw (AT2019php) & 0.087 & $-18.9\pm0.1$ & $3.6\pm0.1$ & $4.5\pm0.3$ & SN\,Ibn \\
19aapfmki (SN2019deh) & 0.05469 & $-19.90\pm0.01$ & $4.38\pm0.03$ & $7.2\pm0.4$ & SN\,Ibn \\
18abskrix & Galactic & $17.78\pm0.02$ & $1.26\pm0.03$ & $2.5\pm0.2$ & CV & Spectroscopic classification \\
18absrffm (AT2018ftw) & Galactic & $16.34\pm0.01$ & $2.60\pm0.03$ & $5.00\pm0.03$ & CV & Spectroscopic classification \\
18abyzkeq & Galactic & $18.32\pm0.10$ & $1.37\pm0.04$ & $0.93\pm0.06$ & CV & AAVSO Name: CSS 151114:224934+375554 \\
18ablxawt & Galactic & $18.31\pm0.04$ & $2.4\pm0.3$ & $5.2\pm0.9$ & Likely flare star & Previous detection in PS1 DR2 at $i=19.4$ \\
19abpwygn & - & $16.74\pm0.01$ & $2.03\pm0.03$ & $1.73\pm0.03$ & Likely flare star & Previous detection in PS1 DR2 at $z=18.75$ \\
18abyjgaa & - & $18.39\pm0.03$ & $0.82\pm0.03$ & $2.05\pm0.08$ & Likely flare star & GALEX source, possible flaring in PS1 DR2 \\
18aasaiyp & 0.104 & $-19.13\pm0.05$ & $1.9\pm0.6$ & $17.1\pm0.6$ & Unknown & \\
18abuvqgo & 0.155 & $-19.93\pm0.05$ & $4.7\pm0.2$ & $9.9\pm0.6$ & Unknown \\
18abydmfv (AT2018hkr) & 0.042 & $-18.66\pm0.03$ & $3.15\pm0.04$ & $7.7\pm2.5$ & Unknown \\
18acepuyx (AT2018kxh) & 0.0711 & $-19.1\pm0.2$ & $1.4\pm0.3$ & $10.8\pm1.2$ & Unknown  \\
19aatoboa (AT2019esf) & 0.0758 & $-18.90\pm0.03$ & $2.41\pm0.03$ & $4.9\pm0.3$ & Unknown \\
19abgbbpx (AT2019leo) & 0.0625 & $-18.83\pm0.03$ & $4.2\pm0.3$ & $>5$ & Unknown \\
19abiyyhd (AT2019lwj) & 0.07 & $-18.11\pm0.05$ & $2.5\pm0.2$ & $4.1\pm0.3$ & Unknown \\
19aaadfcp & 0.08--0.15 & $19.04\pm0.04$ & $2.44\pm0.15$ & $5.86\pm0.15$ & Unknown \\
19aanvhyc (AT2019coi) & 0.056--0.076 & $18.41\pm0.04$ & $4.39\pm0.04$ & $12.1\pm2.1$ & Unknown \\
18abxxeai & & $18.55\pm0.06$ & $1.9\pm0.1$ & $6.0\pm0.8$ & Unknown & `PSF' host in LegacySurvey DR8\\
18acgnwpo & - & $18.90\pm0.05$ & $0.52\pm0.03$ & $6.5\pm0.5$ & Unknown & No clear host \\
19aanqqzb & - & $16.63\pm0.04$ & $1.91\pm0.07$ & $1.2\pm0.1$ & Unknown & No clear host \\
19aaqfdvu & - & $19.02\pm0.06$ & $2.0\pm0.2$ & $1.6\pm0.4$ & Unknown & No clear host \\
19aaxfqyx & - & $18.76\pm0.03$ & $0.98\pm0.03$ & $4.68\pm0.27$ & Unknown & No clear host \\
19abfzfbs & - & $19.36\pm0.17$ & $3.7\pm2.2$ & $13.4\pm3.2$ & Unknown & No clear host \\
\enddata
\end{deluxetable*}

We take eight as an upper limit for the number of transients in ZTF that could fall within the phase-space of Figure~\ref{fig:lum-rise}.
Of these, three peak brighter than 19\,mag,
and four have a peak between 19 and 19.75\,mag.
We now calculate two all-sky rates.
First we assume that the transient peaks at $<19$\,mag, in which case we discard field-nights with a limiting magnitude shallower than 19.75\,mag. Then we assume that the transient peaks at $<19.75$\,mag, in which case we discard field-nights with a limiting magnitude shallower than 20.5\,mag.

Each ZTF field is 47\,deg$^2$, but there is latitude-dependent overlap that has to be taken into account when converting this to a rate per square degrees in the sky.
For the primary grid, a rough estimate of the fill factor is 87.5\%.
For the 1DC survey, the footprint is 10\% smaller than the number of fields multiplied by 47 square degrees.
So, taking fill factor and overlap into account, we estimate a typical area-per-field of 37\,deg$^2$.
So for transients brighter than 19\,mag we have 
$2.5 \times 10^{5}\,\degsq\,\days$
and for transients brighter than 19.75\,mag we have
$1.7 \times 10^{5}\,\degsq\,\days$.
For transients peaking brighter than 19\,mag we have a limiting all-sky rate 

\begin{equation}
    3 \times \frac{41253\,\degsq}{2.5 \times 10^{5}\,\degsq\,\days} \times \frac{365\,\days}{1\,\yr} \approx 180\,\pyr.
\end{equation}

For transients peaking brighter than 19.5\,mag we have a limiting all-sky rate
\begin{equation}
    4 \times \frac{41253\,\degsq}{1.7 \times 10^{5}\,\degsq\,\days} \times \frac{365\,\days}{1\,\yr} \approx 350\,\pyr.
\end{equation}

Now, we use the limiting magnitude to estimate a volumetric rate.
Assuming a transient that peaks at $M=-20$\,mag,
requiring a peak apparent magnitude brighter than 19\,mag
restricts our sensitivity to 400\,Mpc.
So, we find a volumetric rate of $7 \times 10^{-7}\,\pyr\,\pmpccub$.
Requiring a peak apparent magnitude brighter than 19.75\,mag
restricts our sensitivity to 560\,Mpc,
leading to a volumetric rate of $4 \times 10^{-7}\,\pyr\,\pmpccub$.
For reference, we provide rates of core-collapse SNe and GRBs in Table~\ref{tab:rates}.
The rate of events like \name\ appears to be at least two orders of magnitude smaller than the CC SN rate,
and more similar to the rate of GRBs in the local universe.

\begin{deluxetable*}{lrr}[htb!]
\tablecaption{Local ($z=0$) Rates of core-collapse supernovae and GRBs. Approximately 30\% of CC SNe arise from a progenitor stripped of its hydrogen envelope. Among these stripped events, there are roughly equal numbers of IIb, Ib, and Ic events. Of the Ic events, $\sim 10\%$ are ``broad-lined'' with photospheric velocities $\gtrsim$30,000\,km/s. The fraction of Ic-BL SNe with associated GRBs has been estimated to be 1/40 \citep{Graham2016} although the rate is highly uncertain. The fraction of Ic-BL SNe with associated LLGRBs remains uncertain. Note that the rate quoted for LLGRBs does not include a beaming correction.\label{tab:rates}}
\tablehead{\colhead{Class} & \colhead{Rate/Fraction} & \colhead{References}}
\tabletypesize{\normalsize}
\startdata
SN II & $4.47 \pm 1.39 \times 10^{-5} \, \mathrm{yr}^{-1} \, \mathrm{Mpc}^{-3} $ & [1] \\
SN Ibc & $2.58 \pm 0.72 \times 10^{-5} \, \mathrm{yr}^{-1} \, \mathrm{Mpc}^{-3} $ & [1] \\
Frac. of Ibc SN that are Ic & $0.69 \pm 0.09$ & [2,3] \\
Frac. of Ic SN that are Ic-BL & $0.21 \pm 0.05$ & [2,3] \\
LLGRB & $2.3_{-1.9}^{+4.9} \times 10^{-7}\,\pyr\,\pmpccub$ & [4] \\
& $3.3_{-1.8}^{+3.5} \times 10^{-7}\,\pyr\,\pmpccub$ & [5] \\
$\ell$GRB & ${\cal{R}}_\mathrm{obs} = 4.2^{+9.0}_{-4.0} \times 10^{-10} \,\pyr\,\mathrm{Mpc}^{-3}$ & [6] \\
 & $f_b = 0.0019 \pm 0.0003$ & [7] \\
  & $f_b = 0.013 \pm 0.004$ & [8] \\
\enddata
\tablereferences{[1] \citet{Li2011}, [2] \citet{Kelly2012}, [3] \citet{Graham2016}, [4] \citet{Soderberg2006}, [5] \citet{Liang2007}, [6] \citet{Lien2014}, [7] \citet{Frail2001}, [8] \citet{Guetta2005}}
\end{deluxetable*}

\section{Prospects for Detecting X-ray Emission}

Clearly, radio observations are an important avenue of follow-up for transients like \name.
Another valuable avenue is X-ray observations, which were not obtained for \name.
We can estimate what the predicted X-ray luminosity would be from inverse Compton scattering, using the optical and radio luminosities:

\begin{equation}
    \frac{L_X}{L_\mathrm{radio}}=\frac{u_{ph}}{u_B}.
\end{equation}

Taking $L_\mathrm{radio}=10^{40}\,\erg\,\psec$, $u_{ph} = 10^{44}\,\erg\,\psec/(4 \pi R^3 /3)$ where $R=10^{14}\,\cm$,
and $u_B = B^2/8\pi$ where $B=0.6\,$G,
we find $L_X \approx 10^{43}\,\erg\,\psec$.
This is even more luminous than the X-ray emission observed accompanying AT2018cow, 
which had $L_X\approx10^{42}\,\erg\,\psec$ \citep{Rivera2018,Ho2019a,Margutti2019}.
To our knowledge there were no X-ray follow-up observations of DES16X1eho,
while observations of iPTF16asu resulted in an X-ray upper limit of $10^{43}\,\erg\,\psec$.
\citet{Hosseinzadeh2017} report pre-peak UV measurements from \swift\ for iPTF15ul, but to our knowledge X-ray observations have not been reported.
We measured an upper limit of $0.005\,\ct\,\psec$ in a single epoch from the publicly available \swift\ data.
Assuming $n_H=1.7 \times 10^{20}\,\pcmsq$ and a power-law source model with a photon index $\Gamma=2$ we obtain an upper limit on the unabsorbed 0.3--10\,\kev\ luminosity of $2 \times 10^{42}\,\erg\,\psec$.

\section{Summary and Conclusions}
\label{sec:conclusions}

\name\ is distinguished by two key characteristics:
a fast-evolving optical light curve with a hot ($T>40,000\,$K) and featureless thermal spectrum at peak, and a long-lived, fast-fading radio light curve similar to those of jet-powered long-duration GRBs.
The host galaxy underwent a recent starforming episode and has a very high specific star-formation rate, similar to that of extreme SLSN and GRB hosts.
Events like \name\ are rare:
from one year of the ZTF 1DC survey,
we estimate that the rate is at least 2--3 orders of magnitude smaller than the CC SN rate.

Due to the lack of late-time photometry,
we cannot conclude whether the late-time light curve
was powered by the same mechanism as the peak or whether another
mechanism such as nickel decay became dominant,
and we have only tentative evidence for color evolution (cooling) over time.
Furthermore,
we cannot determine whether this source
developed supernova features
and whether it most closely resembles a Ic-BL like iPTF16asu, a continuum with emission lines like the Ibn iPTF15ul or the SN/TDE candidate AT2018cow,
or neither.

Among the fast-luminous optical transients in Table~\ref{tab:literature},
only AT2018cow and SN\,2011kl had detected radio emission.
\name\ thus adds to the very small number of events in the literature established to have fast-blue optical light curves,
as well as a separate fast ejecta component that produces luminous radio emission.
Interestingly, most of the well-studied transients in Table~\ref{fig:lum-rise} are associated with a candidate engine-powered explosion.
AT2018cow had a long-lived central engine that powered a fast ($0.1c$) outflow.
The Koala likely had a central engine that powered an even faster ($>0.38c$) outflow, perhaps a relativistic jet.
iPTF16asu was a Ic-BL SN, and therefore by definition had faster ejecta velocities than ordinary core-collapse supernovae,
although there was no evidence for a jet.
SN\,2011kl had a burst of high-energy emission (GRB\,111209A) and an associated luminous afterglow.
Given the sensitivity of the luminosity to the shock speed (Equation 7),
perhaps this apparent relationship between engine-driven supernovae and luminous fast-luminous optical transients should not be surprising.

At $z=0.27$, \name\ was much more distant than AT2018cow ($z=0.0141$),
but the lesson from \S \ref{sec:radio-obs} and \S \ref{sec:rates} is that we should not be deterred by cosmological distances in pursuing X-ray and radio follow-up observations.
The radio emission from \name\ would be easily detectable by the VLA out to $z=0.5$ (assuming $5\,\ujy$ RMS in half an hour of integration time)
or even out to $z=0.8$ (when it would be $30\,\ujy$).
Assuming a \emph{Swift}/XRT sensitivity limit of $4 \times 10^{-14}\,\erg\,\pcmsq\,\psec$,
the X-ray emission from \name\ may have been on the detection threshold.
For a \emph{Chandra} sensitivity limit an order of magnitude deeper,
this may be on the detection threshold at $z=0.7$.
At these larger distances ($z=0.5,z=0.7$)
the optical $g$-band magnitude would be 21.1 and 22.3 respectively.
This is out of reach for current surveys like ZTF, but standard for LSST.
The false positives in such a search are lower-luminosity explosions (Type IIb, II, and Ibn SNe) and CVs.
These can be ruled out via knowledge of the host redshift (and therefore intrinsic luminosity),
so we emphasize the need for extensive and reliable galaxy-redshift catalogs.

The code used to produce the results described in this paper was written in Python
and is available online in an open-source repository\footnote{https://github.com/annayqho/Koala}.

\appendix
\restartappendixnumbering

\section{Light-curve measurements}
\label{sec:appendix-lc}

To construct Table~\ref{tab:literature},
we used observed bands as close as possible to rest-frame $g$:
$g$-band for $z<0.15$,
$r$-band for $0.15 < z < 0.45$,
$i$-band for $0.45 < z < 0.78$, and
$z$-band for $0.78 < z < 1.0$.
We excluded transients with $z>1.0$.
We measured rise and fade times to 0.75\,mag below peak by linearly interpolating the single-filter light curve, and measured uncertainties using a Monte Carlo with 1000 realizations of the light curve.
Additional notes on each transient are below.

For iPTF15ul ($z=0.066$; \citealt{Hosseinzadeh2017})
the uncertainty on the peak magnitude was dominated by the uncertainty from the host-galaxy extinction estimate.
For AT2018cow ($z=0.0141$; \citealt{Prentice2018,Perley2019cow}) we used the time between the last non-detection and the first detection as an upper limit on the rise time,
although we note that interpolation would give 0.4\,\days, much shorter than 3\,\days.
We also corrected for $0.287\,$mag of Galactic extinction,
which was not applied in Table~3 of \citet{Perley2019cow}.
For a lower limit, we used the $o$-band detection before peak (dominated by $r$-band flux at this epoch), corrected for 0.198\,mag of Galactic extinction.
We assumed $g-r=-0.4\,$mag and $g-i=-0.7\,$mag.

For SN\,2011kl ($z=0.677$) we used column $M_{4556}$ in Table~2 of \citet{Kann2019}.
These values are corrected for rest-frame extinction,
and the contributions from the GRB afterglow and host galaxy have been subtracted.
For SNLS04D4ec ($z=0.593$), SNLS05D2bk ($z=0.699$), and SNLS06D1hc ($z=0.555$) we used the $i$-band light curve from \citet{Arcavi2016} and corrected for Milky Way extinction.

For Dougie ($z=0.19$; \citealt{Vinko2015}) we added an additional 0.1\,mag in quadrature to account for the zero-point uncertainty,
and corrected for $0.031\,$mag of Milky Way extinction.
For iPTF16asu ($z=0.187$; \citealt{Whitesides2017}) 
we could not measure the rise or peak magnitude in rest-frame $g$ because observations in the appropriate filter ($r$) began only 3\,days after peak.
We estimated an upper limit to the peak magnitude by assuming that the $g-r$ color at peak was identical to the $g-r$ color during the first $r$-band measurement. We used the first $r$-band measurement as a lower limit.
For the time from half-max to max, we used the observed $g$-band light curve instead.
We obtained the $i$-band light curve of
DES16X1eho ($z=0.76$; \citealt{Pursiainen2018}) from M. Pursiainen (private communication).

\acknowledgements

A.Y.Q.H would like to thank the NRAO staff for their help with data calibration and imaging, particularly Steve Myers, Aaron Lawson, Drew Medlin, and Emmanuel Momjian. She is grateful for their support and hospitality during her visit to Socorro. She thanks Gregg Hallinan and Brad Cenko for their advice on reducing the radio and X-ray data, respectively,
Jochen Greiner and Iair Arcavi for their assistance in obtaining the afterglow-subtracted light curve of SN\,2011kl,
Miika Pursiainen for sharing light curves of DES fast-luminous transients,
Griffin Hosseinzadeh for useful discussions about iPTF15ul,
Jesper Sollerman and Steve Schulze for carefully reading the manuscript,
and Tony Piro and Ben Margalit for other productive conversations.
This work made use of the IPN master burst list
(ssl.berkeley.edu/ipn3/masterli.html) maintained by Kevin Hurley.
We would like to thank Raffaella Margutti for pointing out a typo in an earlier version of this paper,
and to the anonymous referee for detailed comments that greatly improved the flow and clarity of the paper.

A.Y.Q.H. would like to acknowledge the support of a National Science Foundation Graduate Research Fellowship under Grant No.\,DGE-1144469,
the GROWTH project funded by the National Science Foundation under PIRE Grant No.\,1545949,
and the Heising-Simons Foundation.
P.C. acknowledges the support from the Department of Science and Technology via SwarnaJayanti Fellowship awards (DST/SJF/PSA-01/2014-15).
A.H. is grateful for the support by grants from the Israel Science Foundation, the US-Israel Binational Science Foundation, and the I-CORE Program of the Planning and Budgeting Committee and the Israel Science Foundation. 
This research was funded in part by a grant from the Heising-Simons Foundation and a grant from the Gordon and Betty Moore Foundation through Grant GBMF5076, and benefited from interactions with Daniel Kasen and David Khatami also funded by that grant.
A.A.M.~is funded by the Large Synoptic Survey Telescope Corporation, the
Brinson Foundation, and the Moore Foundation in support of the LSSTC Data
Science Fellowship Program; he also receives support as a CIERA Fellow by the
CIERA Postdoctoral Fellowship Program (Center for Interdisciplinary
Exploration and Research in Astrophysics, Northwestern University).

Based on observations obtained with the Samuel Oschin Telescope 48-inch and the 60-inch Telescope at the Palomar Observatory as part of the Zwicky Transient Facility project. ZTF is supported by the National Science Foundation under Grant No. AST-1440341 and a collaboration including Caltech, IPAC, the Weizmann Institute for Science, the Oskar Klein Center at Stockholm University, the University of Maryland, the University of Washington, Deutsches Elektronen-Synchrotron and Humboldt University, Los Alamos National Laboratories, the TANGO Consortium of Taiwan, the University of Wisconsin at Milwaukee, and Lawrence Berkeley National Laboratories. Operations are conducted by COO, IPAC, and UW.
The National Radio Astronomy Observatory is a facility of the National Science Foundation operated under cooperative agreement by Associated Universities, Inc.
We thank the staff of the GMRT that made these observations possible. GMRT is run by the National Centre for Radio Astrophysics of the Tata Institute of Fundamental Research.
Some of the data presented herein were obtained at the W. M. Keck Observatory, which is operated as a scientific partnership among the California Institute of Technology, the University of California and the National Aeronautics and Space Administration. The Observatory was made possible by the generous financial support of the W. M. Keck Foundation.
The authors wish to recognize and acknowledge the very significant cultural role and reverence that the summit of Maunakea has always had within the indigenous Hawaiian community.  We are most fortunate to have the opportunity to conduct observations from this mountain.

\facilities{Keck:I (LRIS), Hale (DBSP), EVLA, GMRT}

\software{
\code{Astropy} \citep{Astropy2013, Astropy2018},
\code{IPython} \citep{ipython},
\code{matplotlib} \citep{matplotlib},
\code{numpy} \citep{numpy},
\code{scipy} \citep{Virtanen2019},
\code{extinction} \citep{Barbary2016},
\code{LPipe} \citep{Perley2019lpipe},
\code{PYRAF-DBSP} \citep{Bellm2016},
\code{IDLAstro}}

\end{document}